\documentclass[iop]{emulateapj}
\usepackage{subfigure}
\slugcomment{Accepted by \apj}
\shorttitle{NGC2976\_VP}
\shortauthors{Adams \it{et al.}}

\begin{document}
\title{The central dark matter distribution of NGC 2976\altaffilmark{*}}
\author{Joshua J.~Adams\altaffilmark{1,2}, Karl Gebhardt\altaffilmark{1,3}, 
Guillermo A.~Blanc\altaffilmark{1,2}, 
Maximilian H.~Fabricius\altaffilmark{4}, Gary J.~Hill\altaffilmark{5,3}, 
Jeremy D.~Murphy\altaffilmark{1}, Remco C. E.~van den Bosch\altaffilmark{5,6}, 
Glenn van de Ven\altaffilmark{6}}
\altaffiltext{*}{This paper includes data taken at The McDonald Observatory of The University of Texas at Austin.}
\altaffiltext{1}{Department of Astronomy, University of Texas at Austin, 1 University Station C1400, Austin, TX 78712, USA}
\altaffiltext{2}{Observatories of the Carnegie Institution of Washington, 813 Santa Barbara Street, Pasadena, CA 91101, USA}
\altaffiltext{3}{Texas Cosmology Center, University of Texas at Austin, 1 University Station C1400, Austin, TX 78712, USA}
\altaffiltext{4}{Max-Planck Institut f\"{u}r extraterrestrische Physik, Giessenbachstra\ss e, D-85741 Garching bei M\"{u}nchen, Germany}
\altaffiltext{5}{McDonald Observatory, University of Texas at Austin, 1 University Station C1402, Austin, TX 78712, USA}
\altaffiltext{6}{Max-Planck Institut f\"{u}r Astronomie, K\"{o}nigstuhl 17, 69117 Heidelberg, Germany}
\begin{abstract}
We study the mass distribution in the late-type dwarf galaxy NGC~2976 
through stellar kinematics obtained with the VIRUS-P integral-field 
spectrograph and anisotropic Jeans models as a test of cosmological 
simulations and baryonic processes that putatively alter small-scale 
structure. Previous measurements of the H$\alpha$ emission-line kinematics 
have determined that the dark matter halo of NGC 2976 is most consistent 
with a cored density profile. We find that the stellar kinematics are best 
fit with a cuspy halo. Cored dark matter halo fits are only consistent with 
the stellar kinematics if the stellar mass-to-light ratio is significantly 
larger than that derived from stellar population synthesis, while the 
best-fitting cuspy model has no such conflict. The inferred mass distribution 
from a harmonic decomposition of the gaseous kinematics is inconsistent with 
that of the stellar kinematics. This difference is likely due to the gas disk 
not meeting the assumptions that underlie the analysis such as no pressure support, 
a constant kinematic axis, and planar orbits. By relaxing some of these assumptions, 
in particular the form of the kinematic axis with radius, the gas-derived solution 
can be made consistent with the stellar kinematic models. A strong
kinematic twist in the gas of NGC~2976's center suggests caution, and we 
advance the mass model based on the stellar kinematics as more reliable. The
analysis of this first galaxy shows promising evidence that dark matter
halos in late-type dwarfs may in fact be more consistent with cuspy dark
matter distributions than earlier work has claimed.
\end{abstract}


\section{Introduction}
\label{sec_intro}
The observations of kinematics in low surface brightness (LSB)
and dwarf-late type galaxies have stubbornly resisted giving clear
evidence for the cuspy Navarro-Frenk-White (NFW) dark matter (DM) halo
profiles that N-body simulations with $\Lambda$CDM inputs predict \citep{Nava96a}.
Instead,
most LSBs and late type dwarfs suggest cored DM halos \citep[e.g.][]{Oh08,deBl08,Kuzi08,Span08,Oh10} or
the observations
are not constraining enough to rule out cusps \citep{Swat03,Simo05}. 
Some simulations have produced cored DM halos by 
rapidly removing the baryonic disk which causes the 
DM halo to expand to a cored equilibrium \citep{Nava96b}, initializing 
numerical simulations with a primordial bar that forms a 
resonance with and disrupts the cusp \citep{Wein02}, or 
by implementing a supernova feedback recipe in 
high-resolution hydrodynamical simulations \citep{Gove10}. 
This puzzle has also motivated proposals for additional 
particle properties of dark matter beyond the weakly interacting, 
cold paradigm such as collisional dark matter \citep[e.g.][]{Sper00}, 
warm dark matter \citep[e.g.][]{Hoga00}, and 
ultra-light boson fluid models \citep[e.g.][]{Peeb00,Good00,Rind11}. 
Other questions critically rely on knowing the DM 
density structure in galaxies, such as the prospects of DM 
annihilation searches that depend sensitively 
on the true density profile in galaxies \citep[e.g.][]{Diem08}. Most 
of the extant attempts to determine DM radial profiles 
rely on gas as the dynamical tracer. A number of works have
studied nearby disk galaxy kinematics with longslit stellar kinematics 
to infer DM halo profiles \citep{Cors99,Corb07} or 
stellar mass-to-light ratios for isothermal sheet models 
\citep{vdKr84,Bahc84,vdKr86,Bott87,Bott89a,Bott89b,Bott90,Bott91,Bott92,Swat99}, 
but better structure constraints come from 2D spectroscopy 
\citep[e.g.][]{Copi04,Kraj05,Capp06,vdBo08,Weij09,Murp11}. 

NGC~2976 has made one of the cleanest cases for a cored
DM halo via its gaseous kinematics \citep[][hereafter SBLB03]{Simo03}. 
In our first attempt to derive DM 
mass profiles from stellar kinematics we chose NGC~2976 
due to several of its 
properties. (1) NGC~2976 is an SAc dwarf galaxy in the M81 group. 
There are some dynamical indications \citep{Spek07} and 
photometric indications \citep{Mene07} that a weak bar may 
be present, but NGC~2976 is usually given an unbarred 
designation \citep{deVa91}. (2) NGC~2976 
has some dark patches that are likely due to dust, but 
its dust content is modest for its Hubble class and 
distributed rather regularly. 
A full treatment regarding the potential impact of dust on the 
measured kinematics is beyond the scope of this work, but several 
literature estimates of the dust content exist. \citet{Will10} fit 
star-formation 
history models from the broadband colors of resolved stars in 
the Advanced Camera for Surveys Nearby Galaxy Survey Treasury (ANGST) 
\citep{Dalc09} by modelling 0.8 magnitudes of differential 
extinction in the V-band for ages above 100 Myr, 
0.5 magnitudes for younger ages, and a foreground screen of 0.46 magnitudes. 
Our data lie within the ``INNER-1'' region of that work.
\citet{Pres07} use Spitzer 24$\mu$m data to estimate 
A$_B\sim$1.5. The lowest estimates comes from SBLB03 
with A$_B\sim$0.23 based on an 
inclination proscription \citep{Saka00}. Although there is a large range in 
the estimated extinctions, the values from each method are on the low end 
for the late-type dwarf population. (3) NGC~2976 appears to be 
dark matter dominated at r$>$500 pc  according to SBLB03 (although 
\citet{deBl08} disagree), so the impact of stellar 
population synthesis (SPS) mass-to-light ($\Upsilon_*$) uncertainties 
is minimized in this target. (4) The stellar surface brightness 
profiles in many bands are very smooth and indicate a bulgeless 
disk with a small nuclear star cluster and a break to an 
outer disk at $\sim$1.2 kpc (SBLB03).

The distance to NGC~2976 
from the tip of the red giant branch method is 3.56$\pm$0.38 Mpc 
\citep{Kara02}; 
we adopt a distance of 3.45 Mpc and a scale conversion of 16.7 pc arcsec$^{-1}$ 
in this work for consistency with SBLB03. The total mass of NGC~2976 
is estimated to be 3.5$\times$10$^{9}$ M$_{\odot}$ based on the 
inclination-corrected line width of 165 km s$^{-1}$ (SBLB03). The inclination 
is variously estimated as 61\fdg5 \citep{deVa91}, 61\fdg4 (SBLB03), and 
64\fdg5 \citep{deBl08}. The HI heliocentric velocity is 4.0$\pm$2.0 km s$^{-1}$ 
\citep{Stil02b}. The stellar velocity 
dispersion in NGC~2976 has not previously 
been reported. 
Measurement attempts were made in \citet{Ho09}, but the results were unresolved 
in the presence of the best $\sigma_{inst}=$ 42 km s$^{-1}$ instrumental resolution and 
unreported. 
They estimate $\sigma=$36.0$\pm$16.8 km s$^{-1}$ from a correlation for their sample with the 
measured [NII]6583\AA\ line width.

We present observations of
NGC~2976 with the large field-of-view fiber fed Visible Integral
field Replicable Unit Spectrograph Prototype (VIRUS-P)
\citep{Hill08a} to concurrently
measure the gaseous and stellar kinematics, fit mass models, and 
study the dark matter halo profile shape in the context of the 
``core-cusp'' controversy with a collisionless tracer. 
Our data reduction and kinematic measurements are 
described in \S \ref{sec_data}.  
We fit the stellar kinematic data with anisotropic Jeans models in 
\S \ref{sec_Jeans}. We perform fits to our [OII] rotation curve 
and the SBLB03 H$\alpha$ rotation curve in \S \ref{sec_gasfit}. 
In \S \ref{sec_BC}, we investigate the constraints on 
$\Upsilon_*$ through matching SPS models to our spectral data and 
its effect on the mass models. Finally, 
we review our conclusions in \S \ref{sec_concl}. As is 
customary, all values of $\Upsilon_*$ are given in solar 
units for the indicated band with the ``*'' indicating a mass 
exclusively for the stellar component. We adopt an absolute 
solar magnitude in the R-band of M$_{\odot,R}=4.42$ throughout 
this work \citep{Binn98}. 
\section{Observations and data reduction}
\label{sec_data}
Over April 27 through May 1 of 2009, we took 18 hours of science pointings on NGC~2976 with the
VIRUS-P instrument and the 2400 lines mm$^{-1}$ VP2 grating on the McDonald Observatory's 
2.7m Harlan J. Smith
telescope under non-photometric conditions. Transparency was 
continuously monitored by stars in the guider camera's data, which was read out and 
saved every five seconds and used to make relative flux calibrations. The 
seeing ranged from 1\farcs4-3\farcs0 full-width-half-maximum (FWHM). The 
VIRUS-P field-of-view covers $1\farcm61\times1\farcm65$ with fibers projected to on-sky 
diameters of 4\farcs235. VIRUS-P has a one-third
filling factor, so we spread three dithers across the galaxy to
maintain continuous spatial coverage in the middle and to
maximize the spatial extent.
We set the instrument to measure 
3680-4400\AA\ with R=2400, nominally. The spectral data were taken under 1$\times$1 binning, 
yielding four pixels sampling the instrumental FWHM. Individual exposures were of 30 
minute duration. 
We took 20 minute sky nods offset from the galaxy by 10\arcmin\ between the science data frames. 
Our best stellar kinematics
come from around the G-band at 4300\AA\ and its many surrounding, mostly Fe, features. We
also measured the [OII]$\lambda\lambda$3726, 3729 lines and all
Balmer lines higher than and including H$\gamma$. The data were reduced 
with the {\tt vaccine} pipeline described in \citet{Adam10a}. We have 
corrected all observations to the heliocentric frame. The 
instrumental wavelength zeropoint is observed to drift by $\sim$10 km 
s$^{-1}$ over normal swings in nightly operating conditions. We 
track this and correct the 
zeropoint by fitting the 4358.3\AA\ Hg I skyline in every 
nodded sky exposure. Quoted wavelengths 
are not corrected to vacuum conditions.
\subsection{Binning}
\label{sec_bin}
\begin{figure}
\centering
\includegraphics [scale=0.7,angle=0]{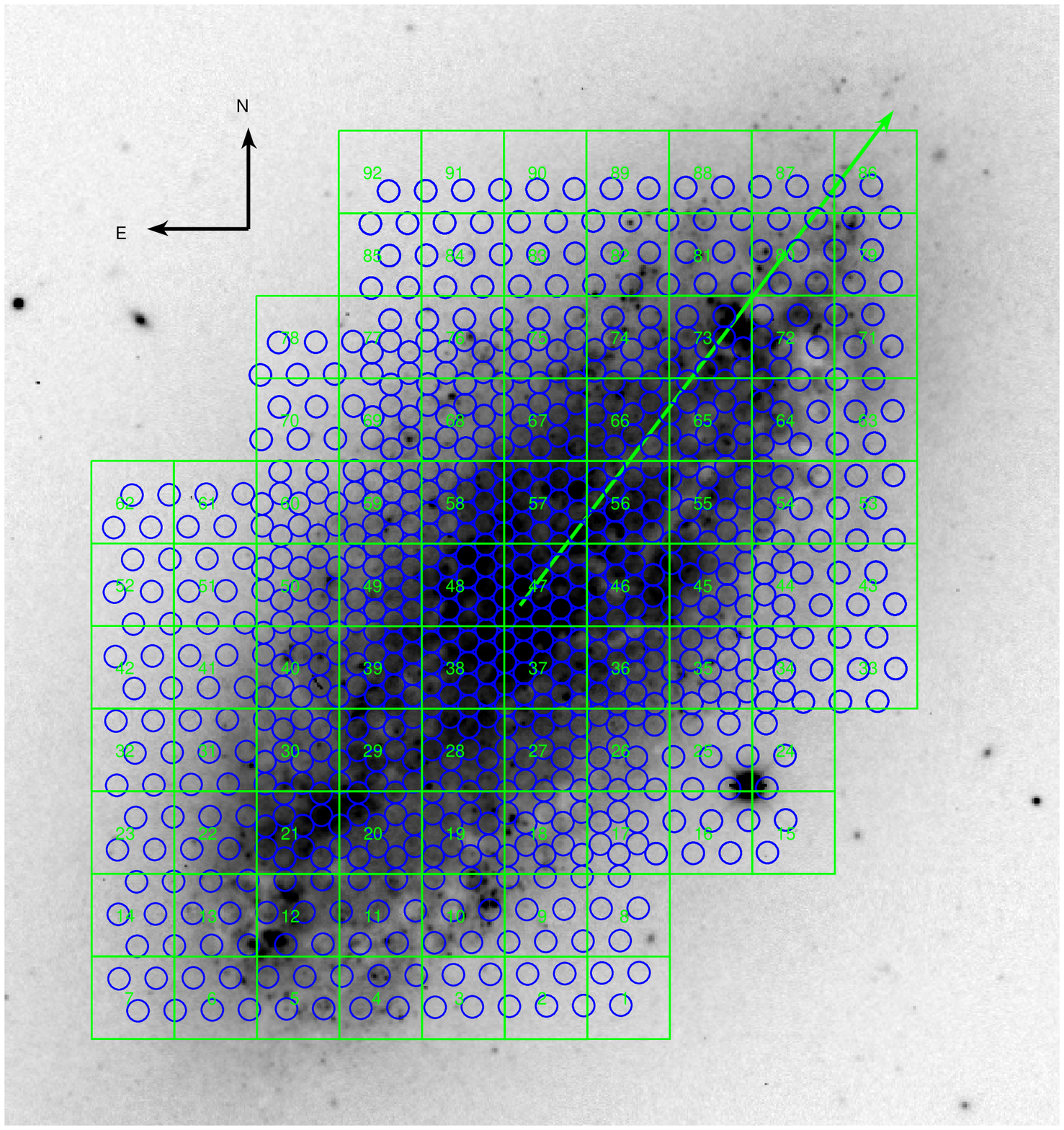}
\caption{The SINGS R-band image of NGC~2976 overlayed with 
the VIRUS-P fiber positions. The numbered squares show 
the spatial bins used in the extraction of the stellar 
kinematics. The arrow indicates the major axis with 
a scale of 120\arcsec\ (2 kpc at our assumed distance).}
\label{fig_layout}
\end{figure}

We have, in total, 738 spectra with signal-to-noise (S/N) ranging up to 
60 per pixel. In order to extract reliable stellar velocity dispersions under 
high S/N conditions, we bin our data as shown in Figure \ref{fig_layout}. 
Constant sized bins were chosen for 
simplicity. The stacks are made with one interpolation to a common, linear 
wavelength scale considering the different wavelength solutions in each 
fiber and the individual heliocentric corrections. Membership in a bin 
was determined solely from each fibers' central position; 
fibers extended over multiple bins were not given partial weights. 
We tested extractions using partial weights, and the results agreed to 
those presented to within the estimated errors. This is expected since 
the square bins are considerably larger than the fibers, and the usual 
fractional area that seeps in and out of the square bin for a 
collection of fibers is only 10\%. 
The spectra were not degraded to have a common instrumental 
resolution prior to 
binning. We made this choice to minimize the covariance between 
spectral channels. Instead, the average instrumental resolution was 
associated to each bin. We compared the kinematics derived from
making resolution-matched stacks and from simply 
degrading the kinematic template spectra to the
average instrumental resolution in each bin. The two methods agreed, 
on average, and only displayed differences within the estimated 
errors. Adaptive binning strategies were not adopted, so there is a significant 
range of S/N in the binned spectra. 
The mean S/N per pixel is 29 with a dispersion of 23. We test 
whether this range in S/N influences our extracted
velocities in \S \ref{sec_stel_kin} and find no impact. 
In the case of NGC~2976, adaptive spatial binning is not ideal and would
lead to the loss of spatial resolution in the outer parts where the contribution of 
$V_{los}$ to $V_{rms}$ is significant.
The analysis of the stellar kinematics is made with these 16\arcsec$\times$16\arcsec bins, while 
the analysis of the gas kinematics is made under the native fiber diameters of 
4\farcs235. 
\subsection{Spectral resolution}

\begin{figure}
\centering
\includegraphics [scale=0.5,angle=-90]{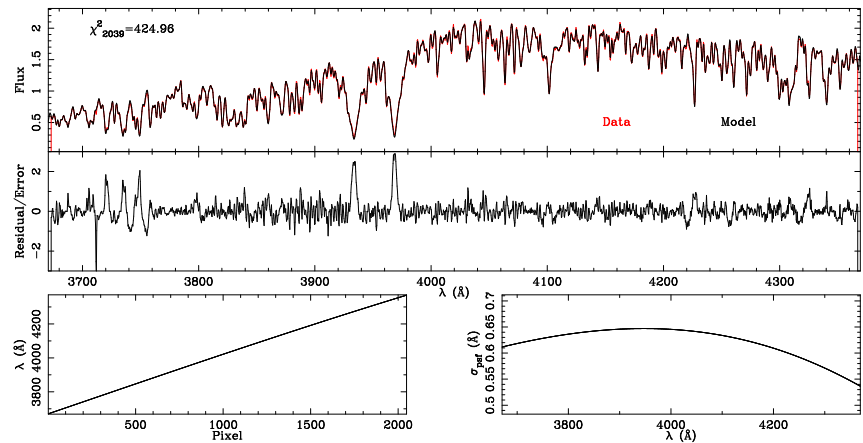}
\caption{The simultaneous wavelength and instrumental 
resolution fit for a particular VIRUS-P fiber. The solutions 
are parameterized by a fourth order polynomial in 
wavelength and a third order polynomial in 
spectral resolution. The fits are made by matching 
a higher resolution solar spectrum to twilight flat frames.}
\label{fig_inst_res}
\end{figure}

\begin{figure}
\centering
\subfigure{\includegraphics [scale=0.5,angle=-90]{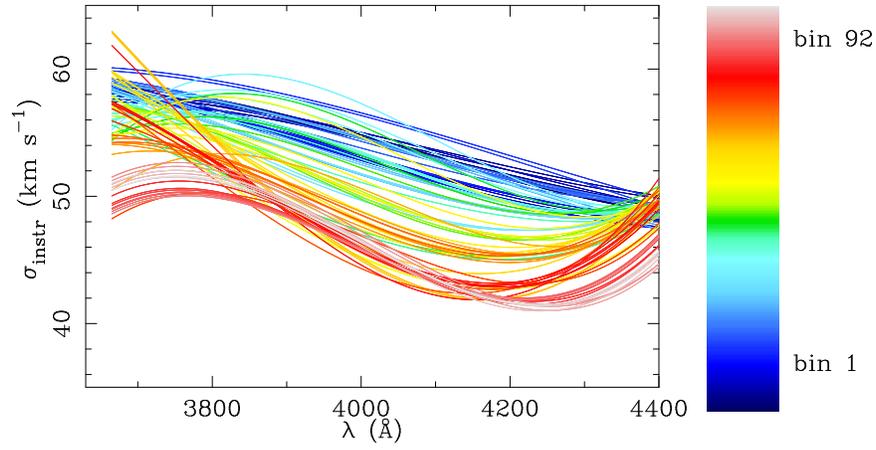}}
\caption{The instrumental resolution in each
kinematic extraction bin.}
\label{fig_fib_res}
\end{figure}

Our observations were taken with an instrumental
dispersion of between 40-60 km/s across different fibers and
wavelengths. We measure this resolution 
to 0.5 km s$^{-1}$ uncertainty in every fiber and wavelength 
with our twilight flats and
high resolution solar spectrum data \citep{Kuru84}. 
Finally, the binning process combines fibers with
different instrumental resolutions, per our dithering scheme.
We average the instrumental resolution for each bin. 
A fit to one fiber is shown in Figure \ref{fig_inst_res}. The 
fit to all bins is summarized in Figure \ref{fig_fib_res}. 
The solutions agree with arc lamp data, although the sparse 
number of available arc lamp lines leads to a less constrained 
solution. We also took spectra of a number of 
template stars \citep{Prug01} to test our instrumental 
resolution and find agreement. We experiment with 
degrading all the data in a bin to the maximum instrumental 
resolution prior to stacking, but we find differences 
only at levels far smaller than the formal errors. 
\label{sec_spec_res}
\subsection{Gaseous kinematics}
We measure the gaseous kinematics through the 
[OII]$\lambda\lambda$3726.032, 3728.815 
doublet. Without binning and in 
each fiber, we simultaneously fit two Gaussian functions over an 18\AA\ 
window. Five parameters are fit 
through a least-squares minization: the intensities of 
each emission line, a constant continuum, the 
radial velocity, and the line width. The 
best fit models are perturbed with the estimated flux uncertainties 
in Monte Carlo realizations to 
generate velocity uncertainties. The median measured intrinsic line width 
is $\sigma$ = 20.4 km s$^{-1}$ with no strong spatial gradients. We 
make models of the circular velocity profile from the line-of-sight velocity  
measurements in \S \ref{sec_gasfit}. 
\subsection{Stellar kinematics extraction}
\label{sec_stel_kin}
We fit the stellar kinematics in each bin with a 
maximum penalized likelihood estimate of the 
Gaussian line-of-sight velocity distribution (LOSVD) in pixel space via 
code described in \citet{Gebh00a}. 
We use the empirical, R~=~10k stellar templates of \citet{Prug01} (ELODIEv3.1) convolved 
to the instrumental resolution of each bin. 
All data and templates are 
normalized prior to convolution by running two boxcars 
over each spectra. The first boxcar has a 40\AA\ width. 
All pixels deviating from the smoothed continuum estimate by 
$>1.5\sigma$ are masked in the second boxcar, which has a 
14\AA\ width. The specifics of this normalization have little impact, as 
judged by varying the normalization parameters, on the 
convolution since both the flux-calibrated VIRUS-P and ELODIE spectra 
hava very shallow slopes over the extracted wavelengths. 

The choice of templates is made from amongst the 1959 available, although 
only a small number are required to describe the data. The membership is 
determined by manually iterating the list to find a local 
miminum in root-mean-square (rms). Several such minima can be found, but the exact
choice is unimportant so long as similar stellar types
are included. We list our chosen, final set 
in Table \ref{tab_kintemp}. A Wolf-Rayet star takes on a significant 
weight. However, its only function over this wavelength range is to 
lower the combined equivalent width of absorption features and is 
degenerate with, say, the A7m star HD003883. The metallicity has 
been modeled by ANGST at [M/H]=-0.12 for this region and by imposing a 
constraint that metallicity grows with time. 
Most of the F and G giants in our template list 
have higher metallicity. However, experiments including the 
template HD148856, a G8III at [Fe/H]=-0.26, do not improve the fit. 

Although we track formal uncertainties, 
we also tally an empirically determined, systematic uncertainty as the rms from the 
best-fit model. The 
two generally agree, expect at high S/N where template 
mismatch becomes evident. All further uncertainties are based 
on the empirical uncertainty, although these may 
overestimate the actual uncertainty. Uncertainties in velocity are 
determined by making Monte Carlo realizations of the 
best-fitting model with simulated noise determined from the 
residuals. 

\begin{figure}
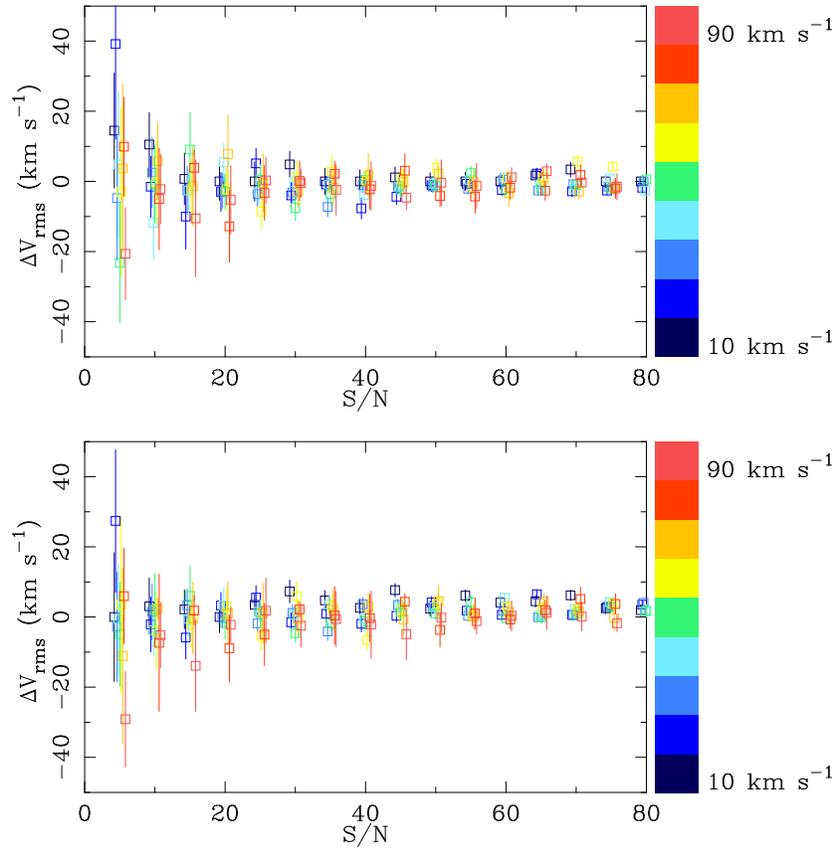

\centering
\subfigure{\includegraphics [scale=0.7,angle=-90]{pgplot_p4_exp1.eps}}\\
\subfigure{\includegraphics [scale=0.7,angle=-90]{pgplot_p4_exp2.eps}}
\caption{The extracted velocity 
dispersions in simulated data. The instrumental
resolution was set to 50 km s$^{-1}$, and the color is coded by
stellar velocity dispersion. \textit{Top}: The 
simulated data came from a subset of the LOSVD template set. 
No systematics are seen down to very low S/N and dispersions. \textit{Bottom}: 
The extracted velocity dispersions in simulated data. The
simulated data come from stars in the ELODIE dataset which are
not contained in our LOSVD template set. This tests for template
mismatch. No systematics are seen down to very low S/N and dispersions.}
\label{fig_disp_test1}
\end{figure}

Tests are run to determine the limits of reliable 
kinematic extraction under a range of S/N and intrinsic 
velocity dispersions. First, a subset of the template spectra 
are combined, then convolved by a simulation dispersion and 
the instrumental resolution, and noise is added. The 
extraction of the velocity 
dispersion is shown in Figure \ref{fig_disp_test1}. The 
errors are accurately estimated, and no systematic effects are 
seen down to S/N$>$5 and $\sigma>10$ km s$^{-1}$. Next, a similar test 
is made to capture the possibility of template mismatch. We 
combine the ELODIE templates for HD000432, HD068380, and 
HD081809 in a 21:53:27\% ratio. The stars are not 
part of our fitting template, but they have similar spectral 
types. The extractions of velocity 
dispersion are shown again in Figure \ref{fig_disp_test1}. 
The errors are marginally larger, but again they are accurately estimated and 
without systematic trends. 

\begin{figure}
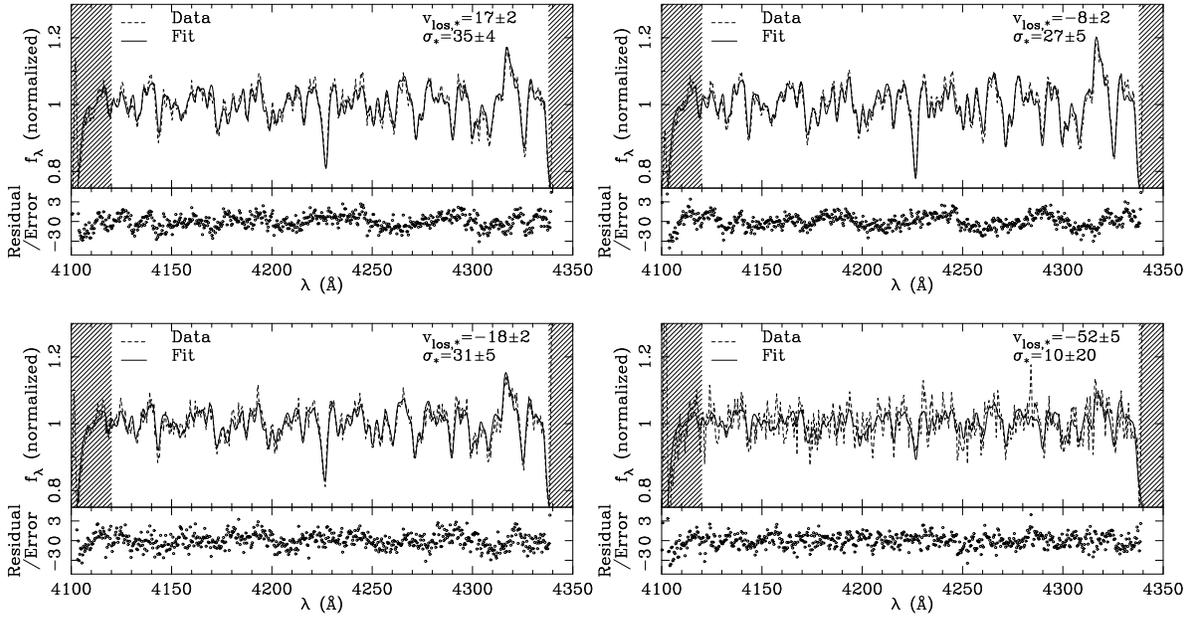

\centering
\subfigure{\includegraphics [scale=0.5,angle=-90]{b56_pgplot_p6.eps}}
\subfigure{\includegraphics [scale=0.5,angle=-90]{b38_pgplot_p6.eps}}\\
\subfigure{\includegraphics [scale=0.5,angle=-90]{b19_pgplot_p6.eps}}
\subfigure{\includegraphics [scale=0.5,angle=-90]{b05_pgplot_p6.eps}}
\caption{Spectra and fits in several bins. The 
chosen bins (56, 38, 19, and 5), represent a range of S/N 
(57, 57, 53, and 26) starting from the 
top left going clockwise. The error normalized 
residuals are shown in the bottom panel.}
\label{fig_stelkin}
\end{figure}

\begin{figure}
\centering
\subfigure{\includegraphics [scale=0.4,angle=0,trim=0 0.3in 0 1.18in,clip=true]{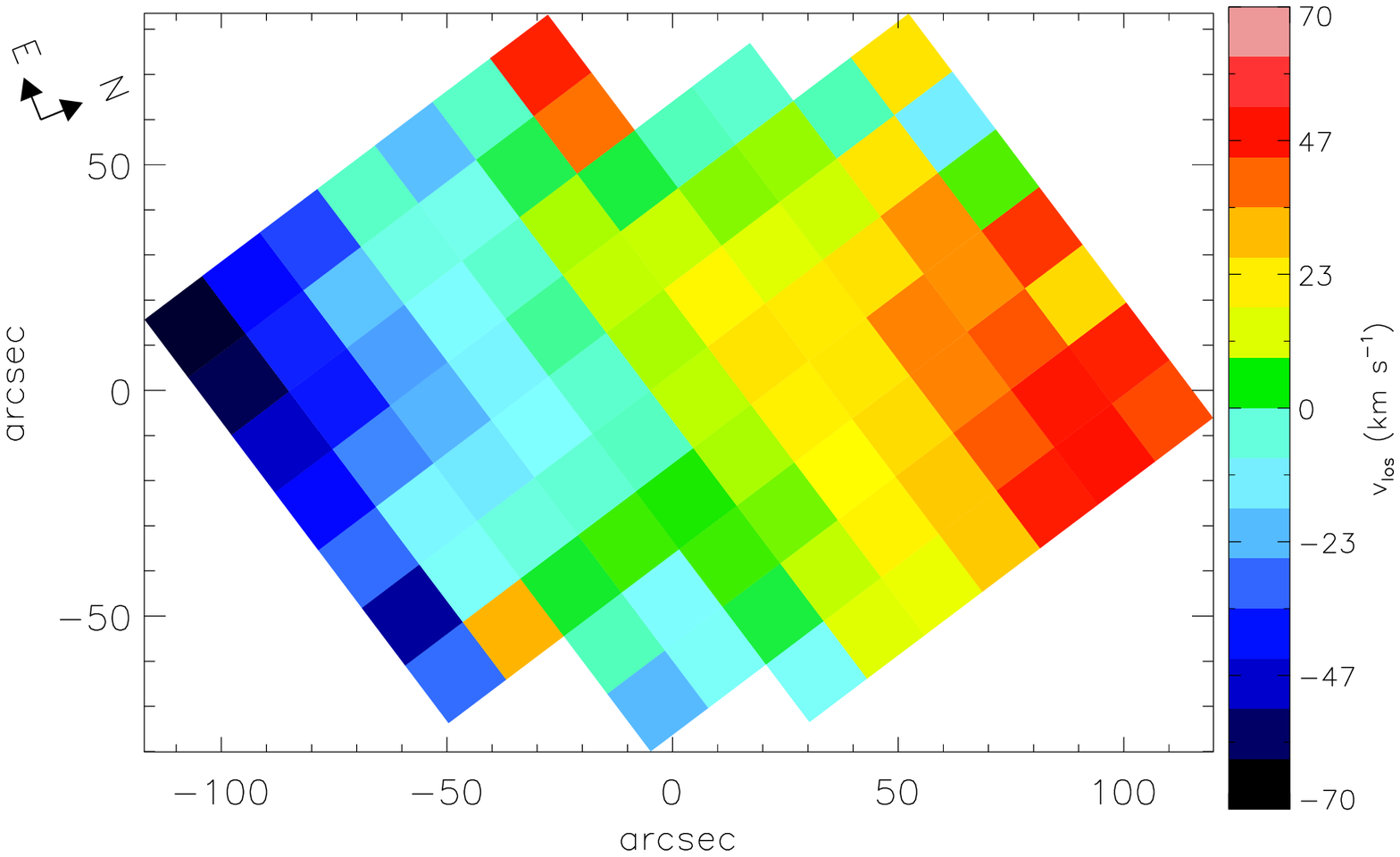}}
\subfigure{\includegraphics [scale=0.4,angle=0,trim=0 0.3in 0 1.18in,clip=true]{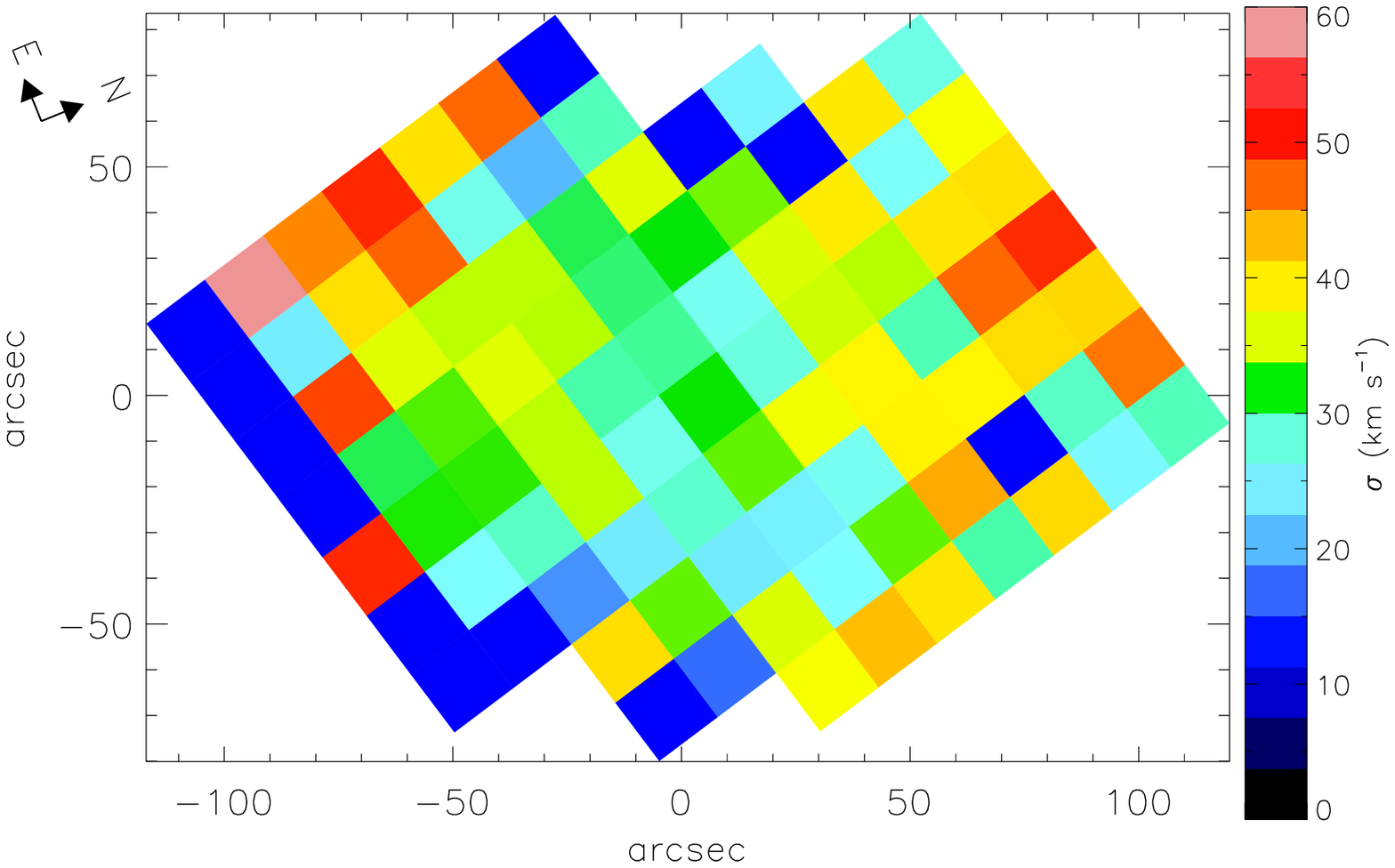}}\\
\subfigure{\includegraphics [scale=0.4,angle=0,trim=0 0.3in 0 1.18in,clip=true]{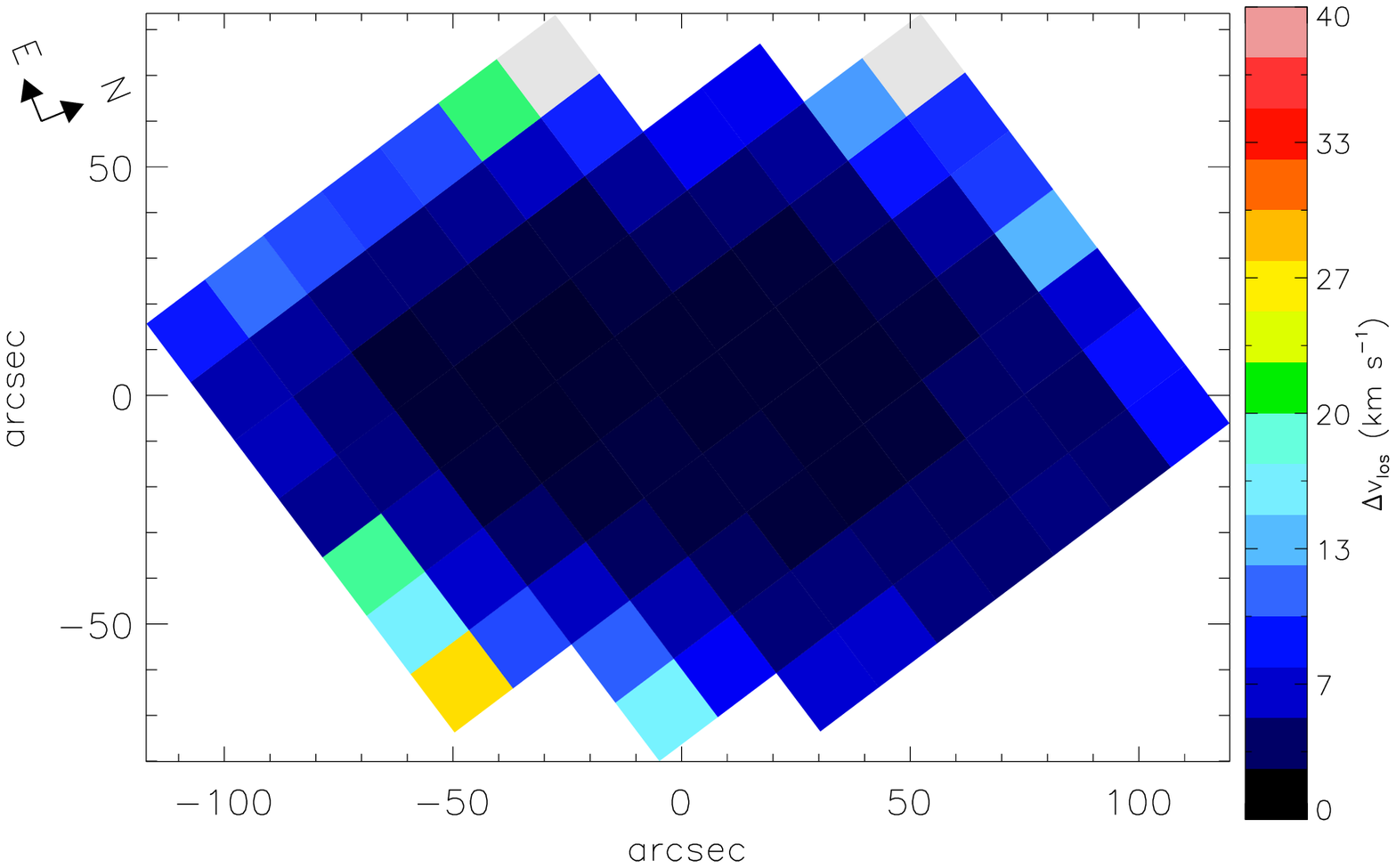}}
\subfigure{\includegraphics [scale=0.4,angle=0,trim=0 0.3in 0 1.18in,clip=true]{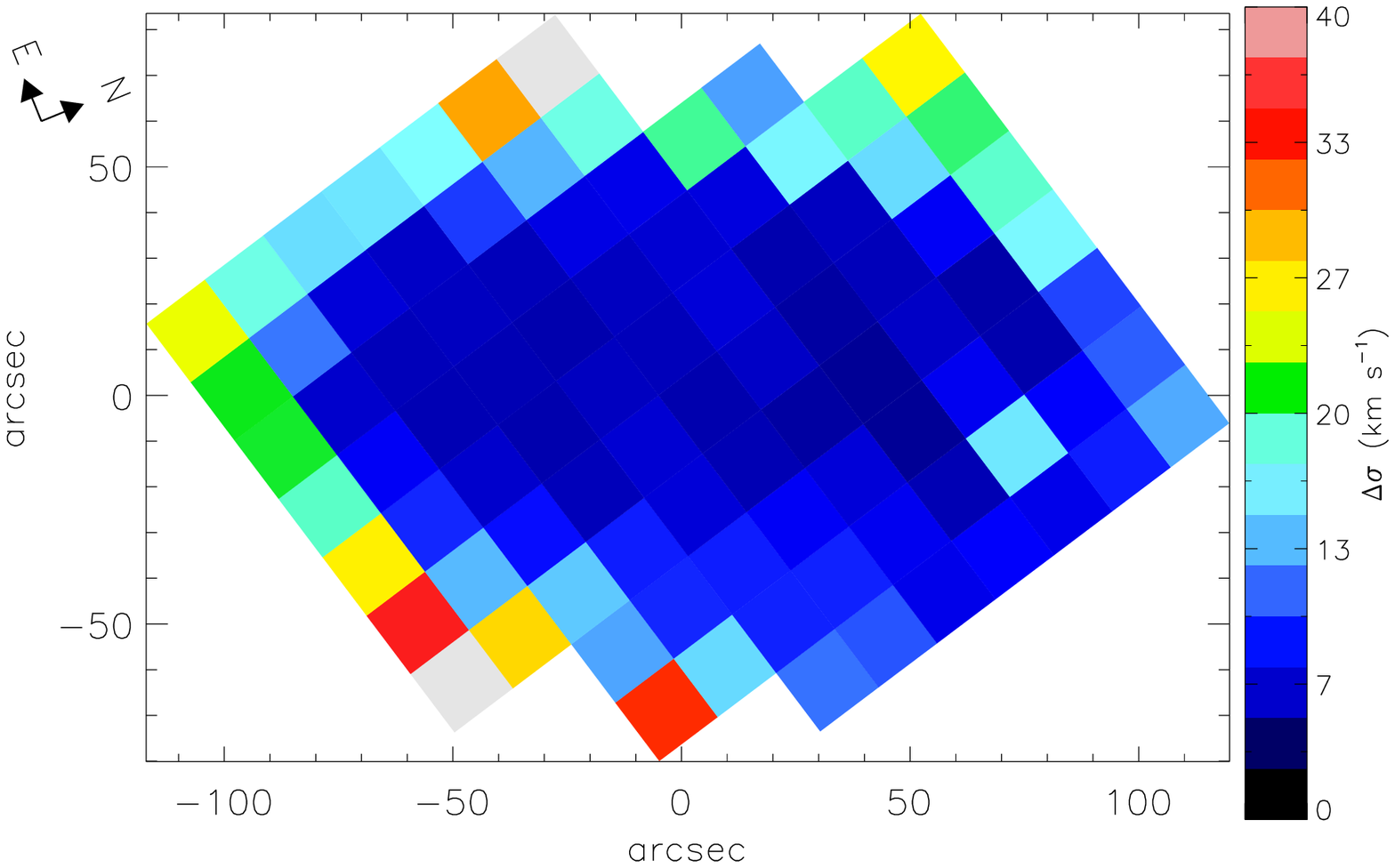}}
\caption{The velocity maps for NGC~2976.\textit{Top left}: The stellar line-of-sight velocity map. 
\textit{Bottom left}: The line-of-sight velocity error map.
\textit{Top right}: The stellar velocity 
dispersion map. The data are consistent with a flat or only 
modestly sloped velocity dispersion profile. 
\textit{Bottom right}: The stellar velocity dispersion error map.
}
\label{fig_velmap}
\end{figure}

Representative spectra and their best-fit models 
are given in Figure \ref{fig_stelkin}. Several additional 
corrections are made. The instrumental resolution 
uncertainty, estimated as 0.5 km s$^{-1}$, is propogated as a random error 
to V$_{rms}$ along with the statistical errors. 
The ELODIE headers quote broadening of the stellar features 
for some stars beyond their R~=~10k resolution; we are not certain 
whether this broadening is physical, such as by binaries, or a spurious artifact. 
The average value for our templates is 
$\sigma=5.2$ km s$^{-1}$. We treat the broadening as physical and 
subtract off the 
ELODIE instrumental resolution 
and the average broadening in quadrature 
from our VIRUS-P instrumental resolution prior to template 
convolution. 
However, the effect is small compared to the final uncertainties. 
Additionally, the systemic velocity is estimated directly from our 
data. The inverse-variance weighted average of our stellar bins is 
4.60 km s$^{-1}$, consistent with an
earlier optical determination of 3$\pm$5 km s$^{-1}$ \citet{deVa91} and 
the HI measurement of 4$\pm$2 km s$^{-1}$ \citet{Stil02b}. 
This is subtracted from the line-of-sight 
velocities before forming V$_{rms}$. 
The line-of-sight velocity and
velocity dispersion for each bin, with uncertainties, 
are given in Table \ref{tab_stelkin}. 
Figure \ref{fig_velmap} shows the maps of line-of-sight velocity, 
velocity dispersion, 
and uncertainties that we measure from the data. The 
final velocity dispersions are not allowed 
to fall below 10 km s$^{-1}$ in order for their error estimates to 
impact the error on V$_{rms}$. This is consistent with the lowest 
velocity dispersions 
that we can reliably extract in simulated data. 

\begin{figure}
\centering
\includegraphics [scale=0.7,angle=-90]{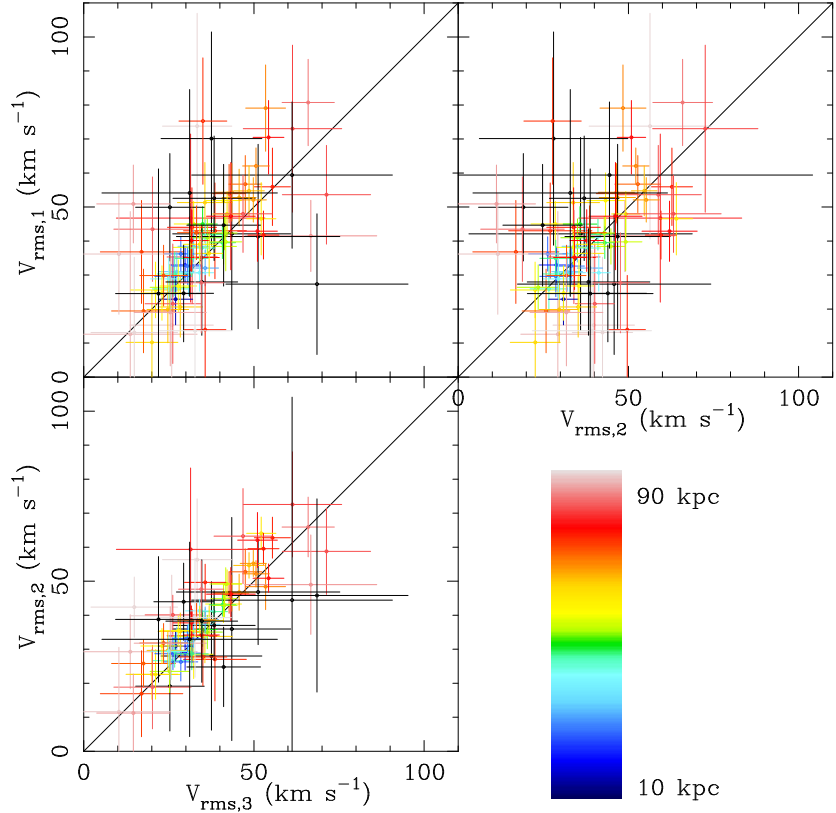}
\caption{The rms velocity as measured in various spectral windows. 
The values are color coded by their distance along the major axis. 
The first, second, and third labels refer to the 3900\AA$<\lambda<$4100\AA, 
4100\AA$<\lambda<$4340\AA, and joint windows respectively. The 
estimates generally agree, especially for the high S/N datapoints which 
drive the modeling. We use the 4100\AA$<\lambda<$4340\AA\ kinematics 
as the preferred value due to the lower errors and the more homogenous 
instrumental resolution compared to the joint range.}
\label{fig_vrms_comp}
\end{figure}

Jeans models make predictions for the  
projected, second velocity moment. For a Gaussian 
kernel, the second-moment is simply V$_{rms}=\sqrt{V_{los}^2+\sigma^2}$. 
The chosen spectral window fit is 4110\AA$<\lambda<$4340\AA\ which includes the G-band, 
a strong Ca absorption line at 4227\AA\ \citep{Wort94}, and a large number of 
weak Fe features. Fits to the 
spectral window 3980\AA$<\lambda<$4100\AA\ give similar measurements, although 
noisier due to the smaller bandpass and less prominent features. We show the consistency 
between spectral regions in Figure \ref{fig_vrms_comp}. 

A careful viewing of 
the velocity dispersion map in Figure \ref{fig_velmap} shows a 
profile that is nearly flat, but perhaps contains higher velocity 
dispersions in the northerly direction. 
This pattern remains whether we remeasure the instrumental resolution within each 
night's data or adopt a resolution fixed with time.
However, the evidence for a 
velocity dispersion gradient is within the noise. 
\section{Stellar population synthesis constraints}
\label{sec_BC}
A loose but independent estimate on $\Upsilon_*$ can be made 
by comparing stellar population models to Spectral Energy Distribution (SED) 
data, either photometrically \citep[e.g.][]{Bell01} or 
spectrophotometrically. Substantial systematic uncertainties 
in, for instance, the initial mass function (IMF) and the properties of 
thermally pulsing asymptotic giant branch (TP-AGB) stars \citep[e.g.][]{Mara05} at 
NIR wavelengths limit the precision of M/L constraints from SED fits. 
Experiments designed to provide 
optimal observational constraints on stellar M/L's by 
applying isothermal sheet relations 
are underway \citep{Herr09,Bers10a,Bers10b}, but the goal of this work 
is instead to model and fit all mass components simultaneously. 
Specific to NGC~2967, SBLB03 found tension 
in the maximal disk value of $\Upsilon_{*,K}<0.09^{+0.15}_{-0.08}$ 
compared to the higher values implied by some SED fits (one of their models has 
$\Upsilon_{*,K}=$2). 

\begin{figure}
\centering
\includegraphics [scale=0.5,angle=-90]{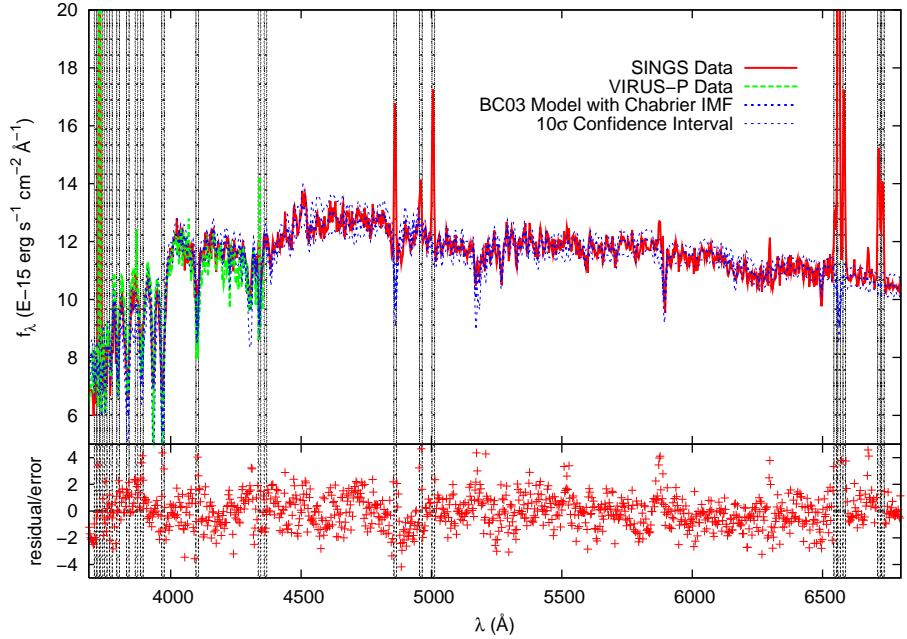}
\caption{Data and fits to synthetic stellar population models. The spectra have 
been corrected for their radial velocity, broadened appropriately, and 
stacked. The SINGS spectrum has poorer resolution than the 
SPS models, so the SPS models are broadened to match. 
The average S/N per pixel over the displayed data is 60. Regions 
with emission lines have been masked in the fit and shaded in grey hatches. 
The best two-component model is shown with confidence intervals. Our 
VIRUS-P data agree well in absolute and relative flux calibration with the 
SINGS spectrum. The SPS fit is made to the SINGS spectrum, preferred by its 
wider bandpass and better coverage of the 4000\AA\ break. The displayed 
best-fit model proscribes $\Upsilon_{*,R}=$ 0.90 with luminosity weighted 
fractions of 0.91:0.09 for a $Z=$ 0.2$Z_{\odot}$, E$_s$(B-V)=0, $\tau_{SFR}=$ 5 Gyr, 12 Gyr age 
model and a $Z=$ 0.2$Z_{\odot}$, E$_s$(B-V) = 0.60, instantaneous burst, 25 Myr age model respectively.
}
\label{fig_stelpop}
\end{figure}

We analyze the stellar population by 
fitting spectra with stellar population synthesis (SPS) models 
from \citet{Bruz03} and the preliminary release of their 2007 version that 
incorporates new TP-AGB values \citep{Bruz07}. We use the Padova 1994 
\citep{Alon93,Bres93,Fago94a,Fago94b,Gira96} and Marigo 2007 
\citep{Mari07} evolutionary tracks, respectively. 
Both Chabrier and Salpeter IMF's are tried, which represent reasonable 
lower and upper bounds on $\Upsilon_*$. To achieve 
robust star formation histories, we use the same 39 templates 
as \citet{Trem04} which entail combinations of 
three metallicities ($Z=0.2,1,2.5 Z_{\odot}$) and a variety of 
star formation histories (instantaneous bursts of age 
0.005, 0.025, 0.10, 0.29, 0.64, 0.90, 1.4, 2.5, 5, and 11 Gyr; 
a 6 Gyr old population under a constant star formation rate, and 
two tau models sampled at a 12 Gyr age with $\tau_{SFR}=5,9$ Gyr). We further 
add a grid of dust extinction with the form of \citet{Calz00} over 21 
values of E$_s$(B-V) uniformly spaced from 0 to 1. We resample our 
spectra to the rest frame, convolve the templates to match the 8\AA\ FWHM 
SINGS instrumental resolution, and 
mask out windows 4\AA\ wide around each of the 
Balmer lines, [OII], and [NeIII]3869. 
Repeated observations of spectrophotometric 
standards stars with VIRUS-P have shown the relative flux calibration 
to be accurate to better than 10\% \citep{Adam10a}. The 
match to the SINGS 20\arcsec$\times$20\arcsec drift-scan 
spectrum is excellent as shown in Figure \ref{fig_stelpop}. 
Due to the larger bandpass, we quote values by fitting to the 
SINGS spectrum. Similar, but less constrained, values 
result from the VIRUS-P data. The templates' 
normalizations are fit through least-square minimization, first 
individually and then in all 334,971 two-component combinations 
from the metallicity, star formation history, and dust grids. The 
relative probability of each model is calculated as 
$\exp(-\chi^2/2)$ \citep[e.g.][]{Kauf03}. All quoted $\Upsilon_*$ 
values include the effects of dust. 
The 1$\sigma$ confidence intervals centered on the highest probability 
$\Upsilon_{*,R}$ are 0.63$\pm$0.39, 1.23$\pm$0.52, 
1.04$\pm$0.23, and 1.42$\pm$0.42 for the BC03/Chabrier, BC03/Salpeter, 
CB07/Chabrier, and CB07/Salpeter models respectively. 
Given the tight ranges from the statistical 
errors alone, the M/L uncertainty is dominated by systematic 
uncertainties. We consider $\Upsilon_{*,R}=$1.1$\pm$0.8, the mean 
and the symmetric uncertainty that encompasses the 1$\sigma$ 
confidence intervals for all four estimates, as the best 
spectrophotometric limit; therefore we 
present mass models with $\Upsilon_{*,R}$ both freely fit and 
penalized by this conservative confidence interval 
in \S \ref{sec_stelfit} and \ref{sec_gasfit}.

\begin{figure}
\centering
\includegraphics [scale=0.5,angle=-90]{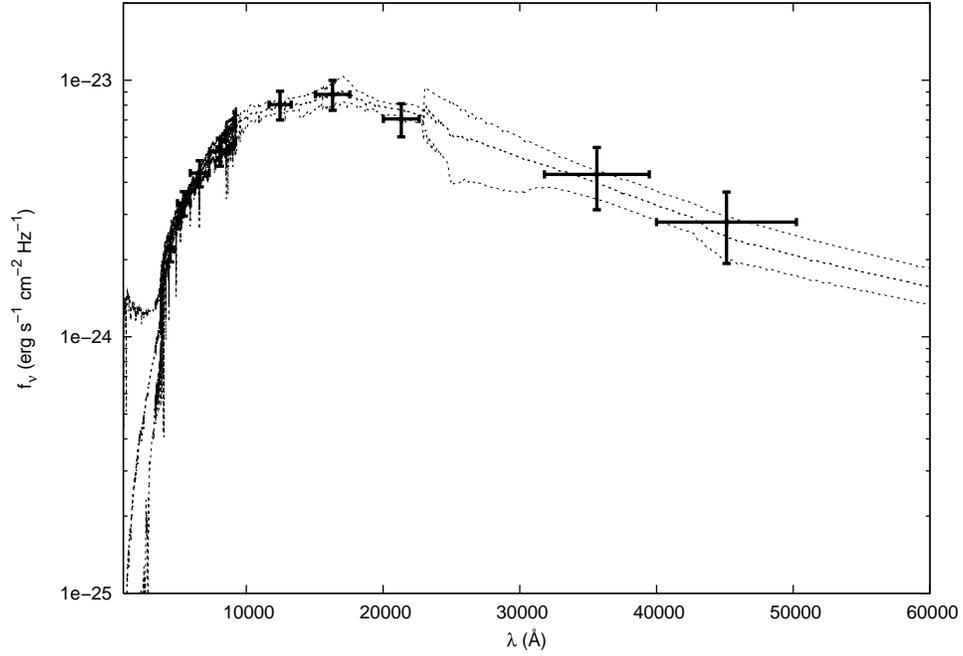}
\caption{The stellar population fits to the optical and NIR data.
The fits predict mass-to-light ratios that are consistent with, but
more loosely constrained, than the fit to the SINGS spectrum. The 
mass modelling of \citet{deBl08} uses the 3.6$\mu$m photometry and a 
color relation derived from \citet{Oh08} for a value of $\Upsilon_{*,3.6\mu m}=$ 0.66 
and claims that NGC~2976 is baryon-dominated. 
Note that this relation was based on stellar populations aged to 12 Gyr 
and does not match the observed K-[3.6 $\mu$m] color of NGC~2976. 
Our models, specifically fit to the spectrophotometry of 
NGC~2976, disagree. We show the best-fit model and confidence interval with 
various two-component SPS models. The best-fit model shown proscribes $\Upsilon_{*,R}=$ 0.64 
and $\Upsilon_{*,3.6\mu m}=$ 0.18 with luminosity 
weighted fractions of 0.67:0.33 for a $Z=$ 0.2$Z_{\odot}$, E$_s$(B-V)=0.40, instantaneous burst, 100 Myr age 
model and a $Z=$ 0.2$Z_{\odot}$, E$_s$(B-V)=0, instantaneous burst, 2.5 Gyr age model, respectively. 
}
\label{fig_BB_SED}
\end{figure}

Finally, we investigate the 
$\Upsilon_*$ limits enabled by broad-band photometry. We use the 
optical-through-2MASS datapoints of SBLB03 by assuming 10$\%$ errors 
and two IRAC datapoints of 
\citet{Dale07} for NGC~2976 without any aperture corrections as 
shown in Figure \ref{fig_BB_SED}. The same stellar population models are 
fit through the EAZY package \citep{Bram08}. The best-fit value 
of $\Upsilon_*$ is consistent with our spectral fits. However, it 
is starkly inconsistent with the SPS value used by \citet{deBl08} 
and calibrated in \citet{Oh08} that renders NGC~2976 to be 
dominated by the baryonic mass. The relation derived there is 
based on SPS models that have aged 12 Gyr and does not 
match the observed colors of NGC~2976. Our fit to the 
SED proscribes $\Upsilon_{*,R}=$ 0.64
and $\Upsilon_{*,3.6\mu m}=$ 0.18 while the \citet{deBl08} 
model proscribes $\Upsilon_{*,3.6\mu m}=$ 0.66. 
\section{Jeans models}
\label{sec_Jeans}
The minimally necessary components to the NGC~2976 mass model are 
a stellar disk with a spatially uniform mass-to-light ratio, 
atomic hydrogen, and a 
dark matter halo under a power-law parameterization. We use 
extant photometry to infer the distribution of the first two, and 
we use kinematic measurements and Jeans model solutions to infer 
the latter.
\subsection{HI deprojection}
We use the robust weighting (${\cal R} = +0.5$), 
zeroth moment map of 21cm atomic hydrogen 
in NGC~2976 from The HI Nearby Galaxy Survey 
\citep[THINGS,][]{Walt08} to characterize the 
HI mass model. The HI distribution 
is highly clumped around the two off-center 
star forming complexes and unlikely to be 
axisymmetric. Nevertheless, we perform 
Multi-Gaussian Expansion \citep[MGE,][]{Capp02} fits to the 
data, subject to the axisymmetric limitations of this 
work's modelling and calculate a resultant 
circular velocity profile \citep[Appendix A,][]{Capp02}. 
SBLB03 have shown that 
HI is a dynamically somewhat important component at r $>80$\arcsec, 
but that H$_2 $ (from CO measurements) is a minor contributor to the 
gravitational potential at all radii. Therefore, we neglect 
the molecular component. 
SBLB03 fit the HI from an older dataset \citep{Stil02a} and by 
assuming an infinitely thin disk; they present a 
circular velocity profile that is 
in general a factor of two times larger than our derived 
values. We apply the MGE mass model to the fits of \S \ref{sec_stelfit} 
and \S \ref{sec_gasfit}, although 
the presence of the HI component does not strongly influence our final results. The contribution 
from HI to the circular velocity is given in Table \ref{tab_gas_rot}.
\subsection{Stellar deprojection}
\label{sec_stel_depr}
\begin{figure}
\centering
\subfigure{\includegraphics [scale=0.4,angle=0]{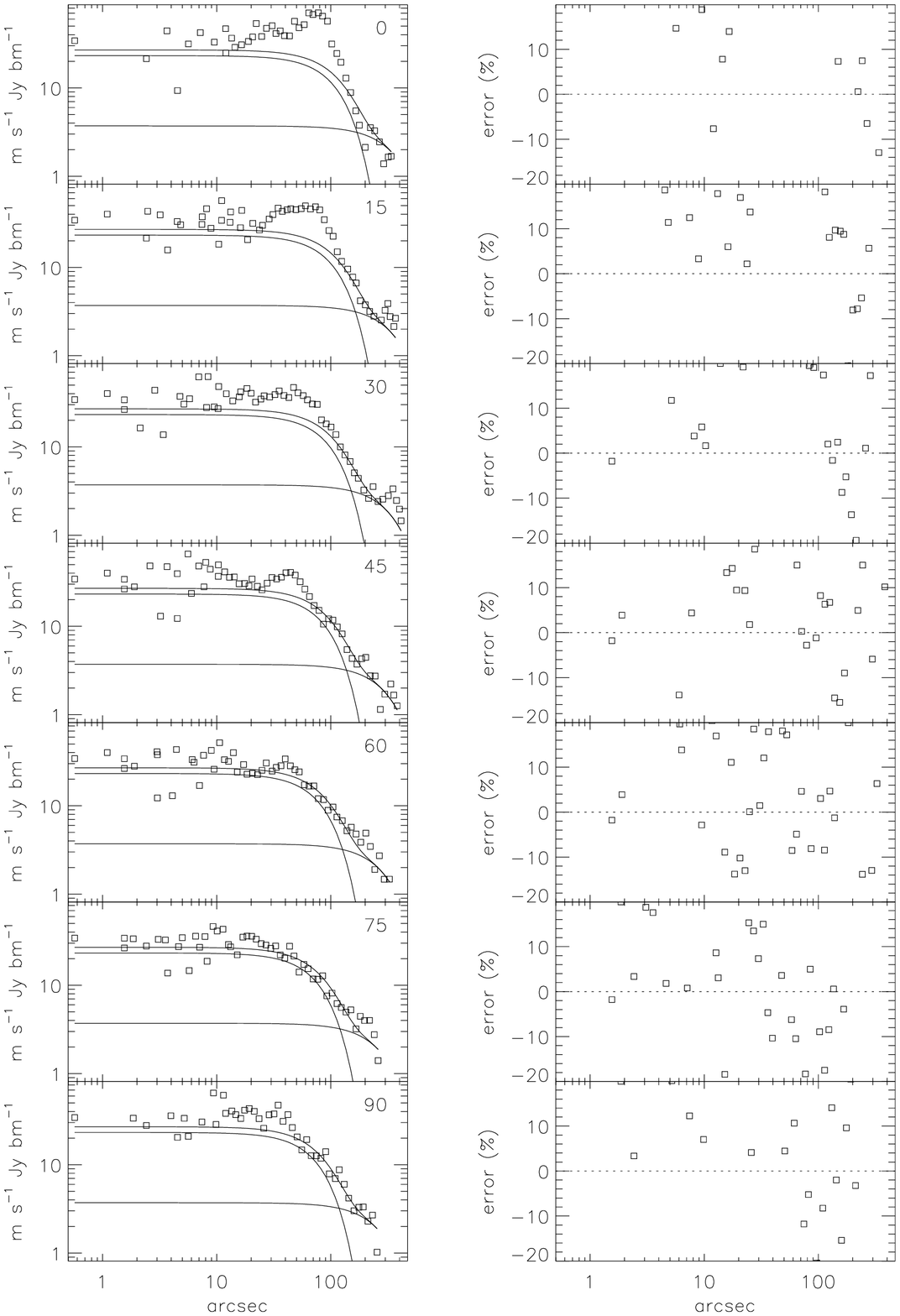}}
\subfigure{\includegraphics [scale=0.4,angle=0]{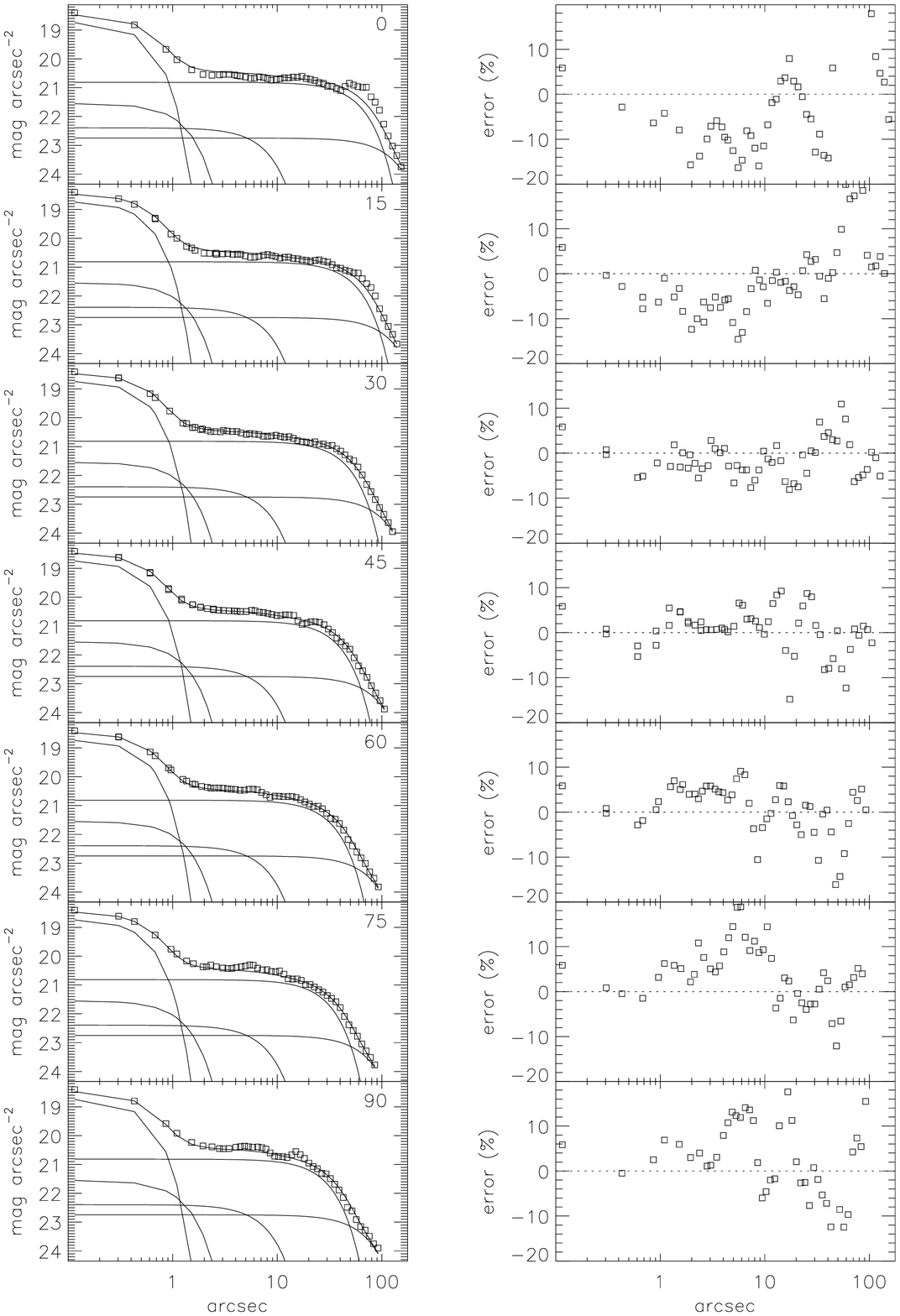}}
\caption{The sector photometry of NGC~2976, folded into one quadrant, with the 
Multi-Gaussian Expansion (MGE) fit also shown. The values in the upper right 
of the profile plot indicate the angle from the major axis, in degrees, 
for each displayed cut. \textit{Left}: 
The HI distribution is strongly asymmetric and requires 
some negative mass components in the model's central regions for a 
quality fit. To maintain a physical density, we have instead made a 
poorer fit with only positive components. However, 
the total HI mass is significantly smaller than the stellar and 
DM mass, and its inclusion or exclusion does not strongly alter the 
DM halo fits.
\textit{Right}: The R-band data and fit. The nuclear stellar
cluster and the break between inner and outer disk near
r$\sim$70\arcsec\ on the major axis and
r$\sim$25\arcsec\ on the minor axis
are captured in the fit. The two star forming regions near
r$\sim$70\arcsec\ on the major axis remain as
residuals since they are asymmetric.}
\label{fig_HI_mass}
\end{figure}

We use the R-band image taken at the Kitt Peak National Observatory's 2.1m 
telescope from the Spitzer Infrared Nearby Galaxies Survey \citep[SINGS,][]{Kenn03} to 
model the stellar mass distribution. The MGE fit is shown in Figure \ref{fig_HI_mass} 
with the terms listed in Table \ref{tab_MGE}. 
The MGE model fits both the nuclear star cluster and the 
inner and outer disks well. The inferred vertical-to-radial
scale length is 1:8 at a nominal $i=63\arcdeg$ over most of the radial range, 
although it is near-circular for the nuclear star cluster.
Strong color gradients are not seen in NGC~2976 (SBLB03), so we limit our 
analysis to a single value of $\Upsilon_{*,R}$ for all components. There is 
some debate as to the best filter to use for accurate recovery of stellar mass. 
Thermally pulsing asymptotic giant branch stars and polycyclic aromatic 
hydrocarbons disfavor the NIR. Dust extinction, varieties of star 
formation history, and nebular 
emission disfavor the optical. The studies on NGC~2976 of SBLB03 and 
\citet{deBl08} used K$_s$ and Spitzer 3.6 $\mu$m data, respectively. 
\citet{Port04} advocates i-band photometry and \citet{Zibe09} calibrates for combinations 
of i and H band photometry. We chose to use the available R-band image 
primarily for its depth and resolution in this work. However, that our 
fits to the spectral energy distribution are consistent across multiple bandpasses (\S \ref{sec_BC}) 
means this choice is unimportant for the analysis of NGC~2976. 

\subsection{Best-fitting dark matter halo}
\label{sec_bestfit}
We use the Jeans Anisotropic MGE modelling package \citep[JAM][]{Capp08} to 
fit the binned stellar kinematic field in NGC~2976 in lieu of 
more computationally intensive Schwarzschild modelling \citep{Schw79}. The fits are made to the 
projected, second-moment velocity ($V_{rms}=\sqrt{V_{los}^2+\sigma^2}$). We 
make models assuming a single anisotropy parameter ($\beta_z=1-(\sigma_z/\sigma_R)^2$) and a 
spatially constant $\Upsilon_{*}$. The spherical DM halo's radial profile 
is approximated as a power law 
\begin{equation}
\label{eq_pl}
\rho(r) = \rho_0\times (r/1\mbox{ pc})^{-\alpha}
\end{equation}
while the NFW 
function has the form of 
\begin{equation}
\label{eq_NFW}
\rho(r) = \frac{\rho_i}{(r/r_s)(1+r/r_s)^2}
\end{equation}
with $r_s$ being a scale radius and 
$\rho_i$ being a density related to the critical density and the halo overdensity. 

The power-law approximiation is justified since the 
core radius of NGC~2976 is likely to lie at 
$r_s \sim 2.5$ kpc (SBLB03, Appendix B), and 
our stellar data do not extend into the asymptotic region 
of the rotation curve. This choice also aids comparison to previous work 
whereas SBLB03 also used this parameterization and 
most models for NGC~2976 in \citet{deBl08} minimized to solutions 
with lower limits on radial scale parameters that then reduce to the 
power-law form. We place the JAM code within a non-linear least-squares minimzation 
package (\texttt{MPFITv.1.64}\footnote[1]{
\texttt{http://cow.physics.wisc.edu/{\raise.17ex\hbox{$\scriptstyle\sim$}}craigm/idl/idl.html}}) 
to reach our optimal solutions. 

\begin{figure}
\centering
\subfigure{\includegraphics [scale=0.3,angle=0,,trim=0 0.3in 0 1.18in,clip=true]{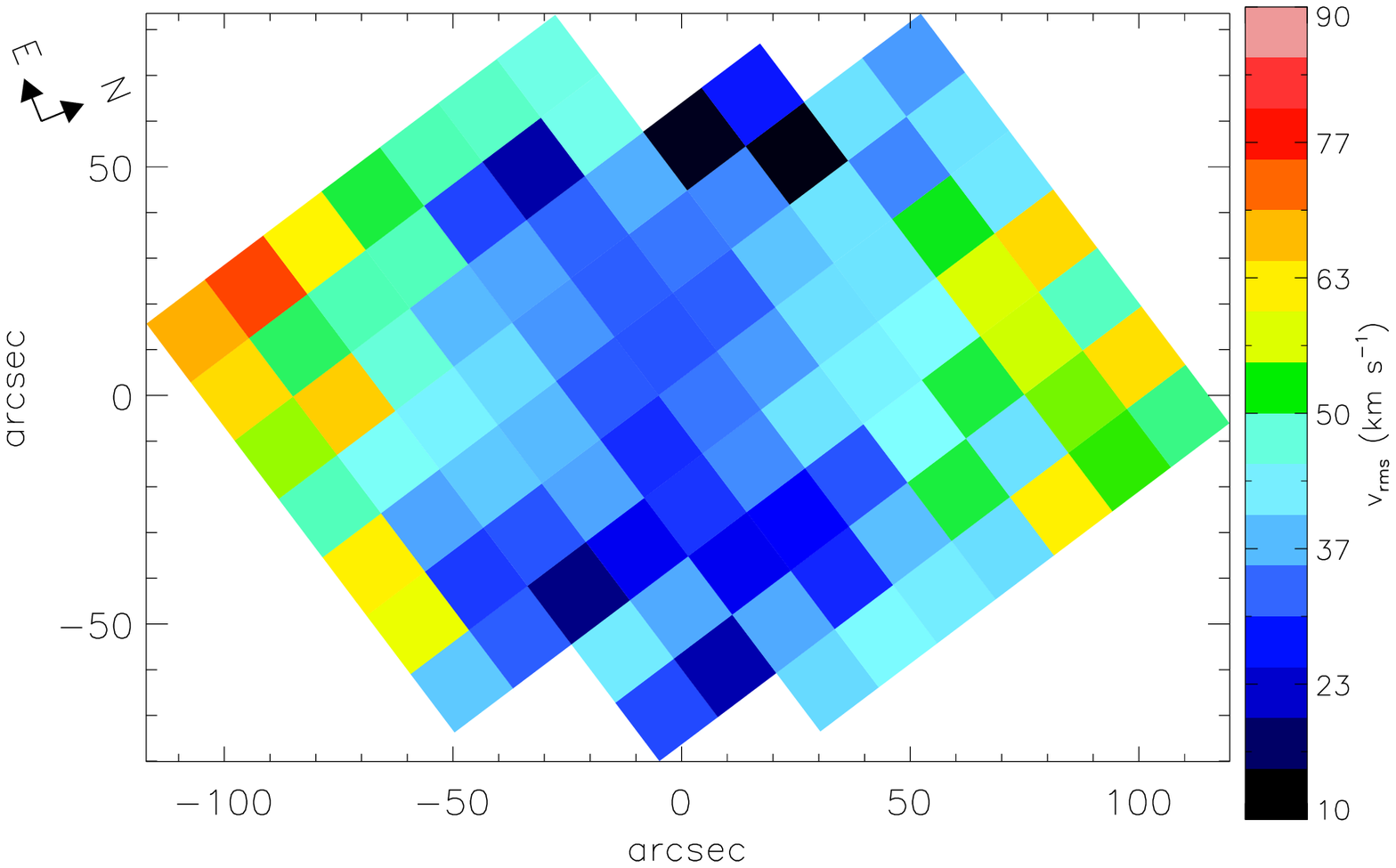}}
\subfigure{\includegraphics [scale=0.3,angle=0,trim=0 0.3in 0 1.18in,clip=true]{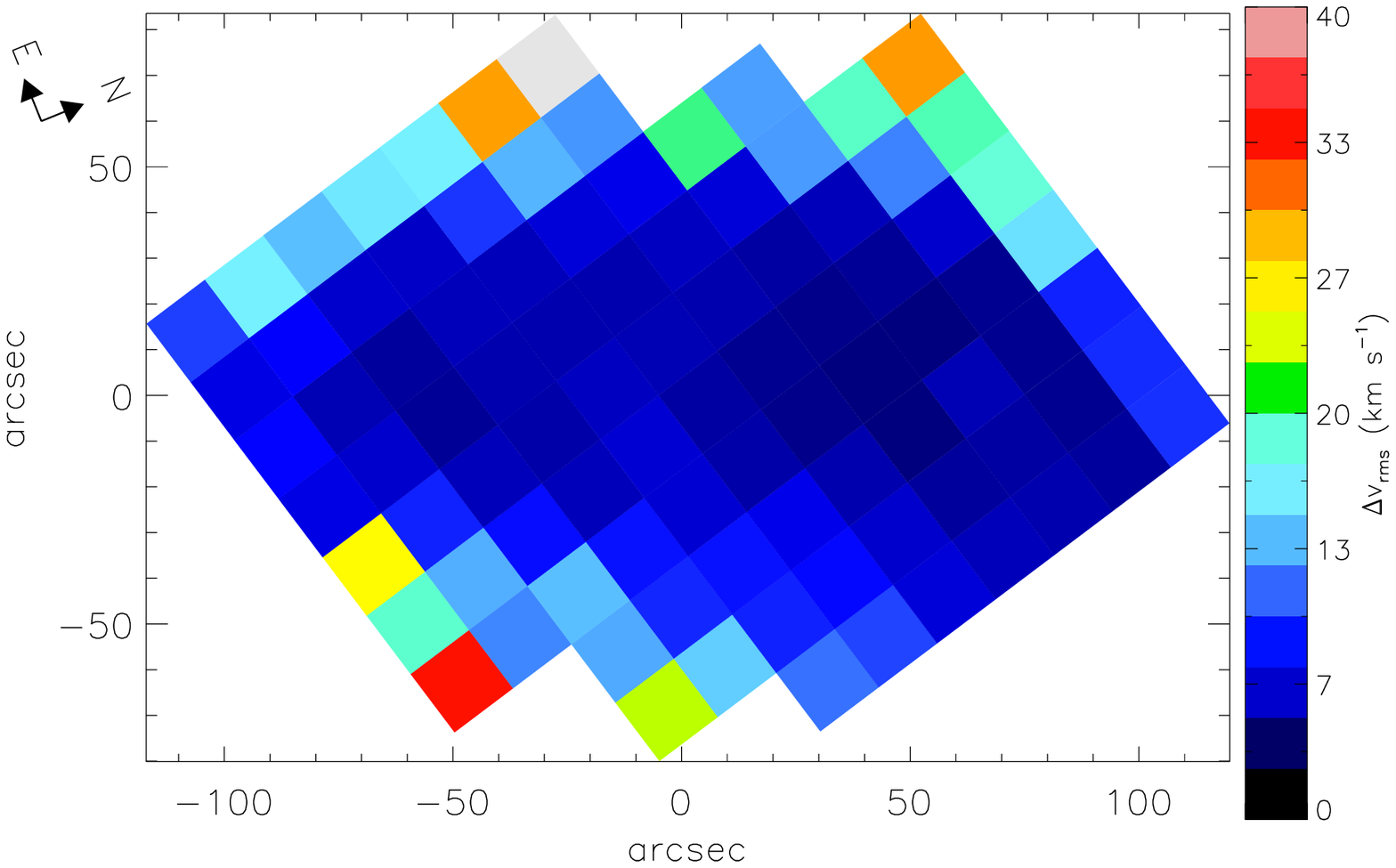}}\\
\subfigure{\includegraphics [scale=0.3,angle=0,trim=0 0.3in 0 1.18in,clip=true]{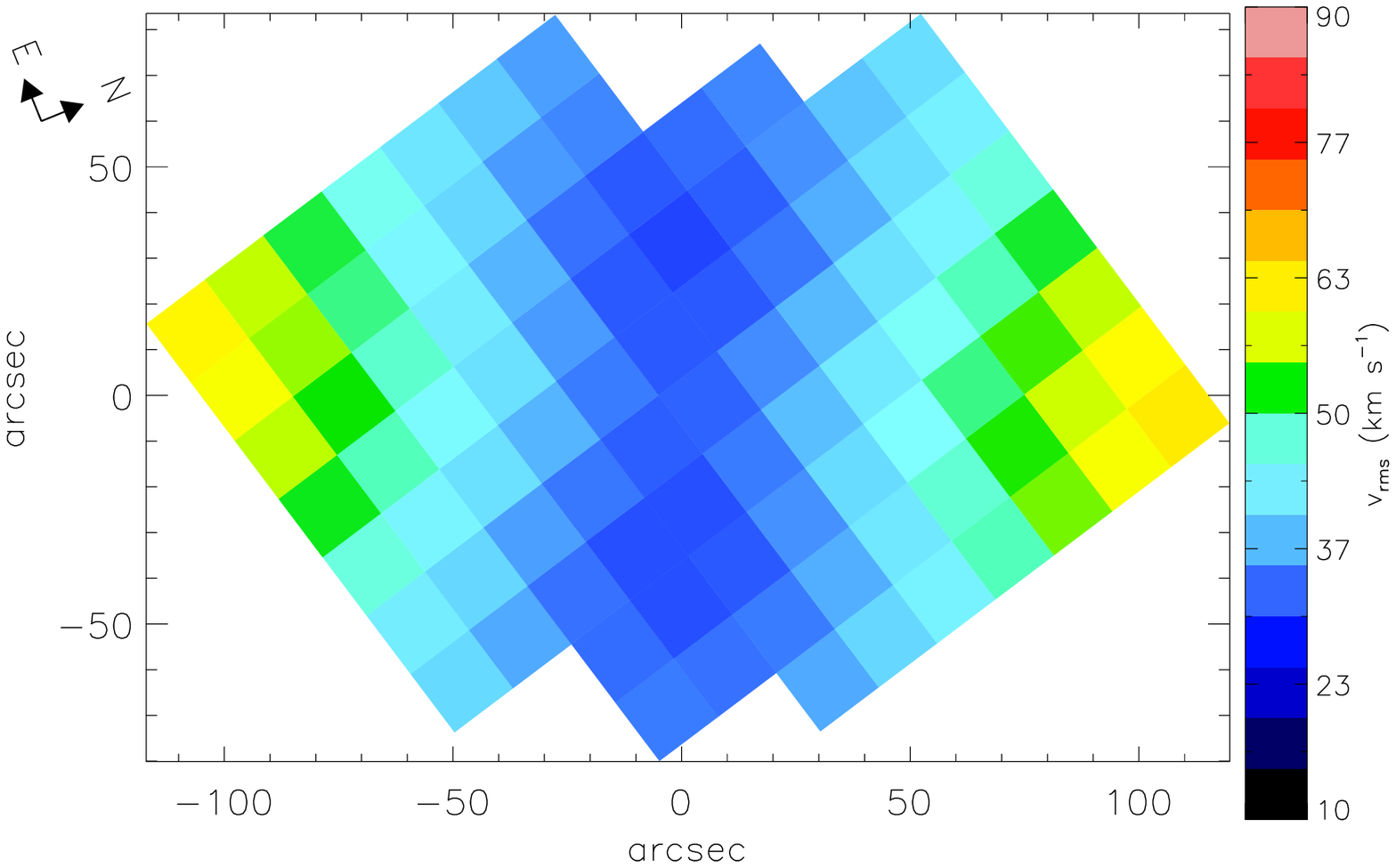}}
\subfigure{\includegraphics [scale=0.3,angle=0,trim=0 0.3in 0 1.18in,clip=true]{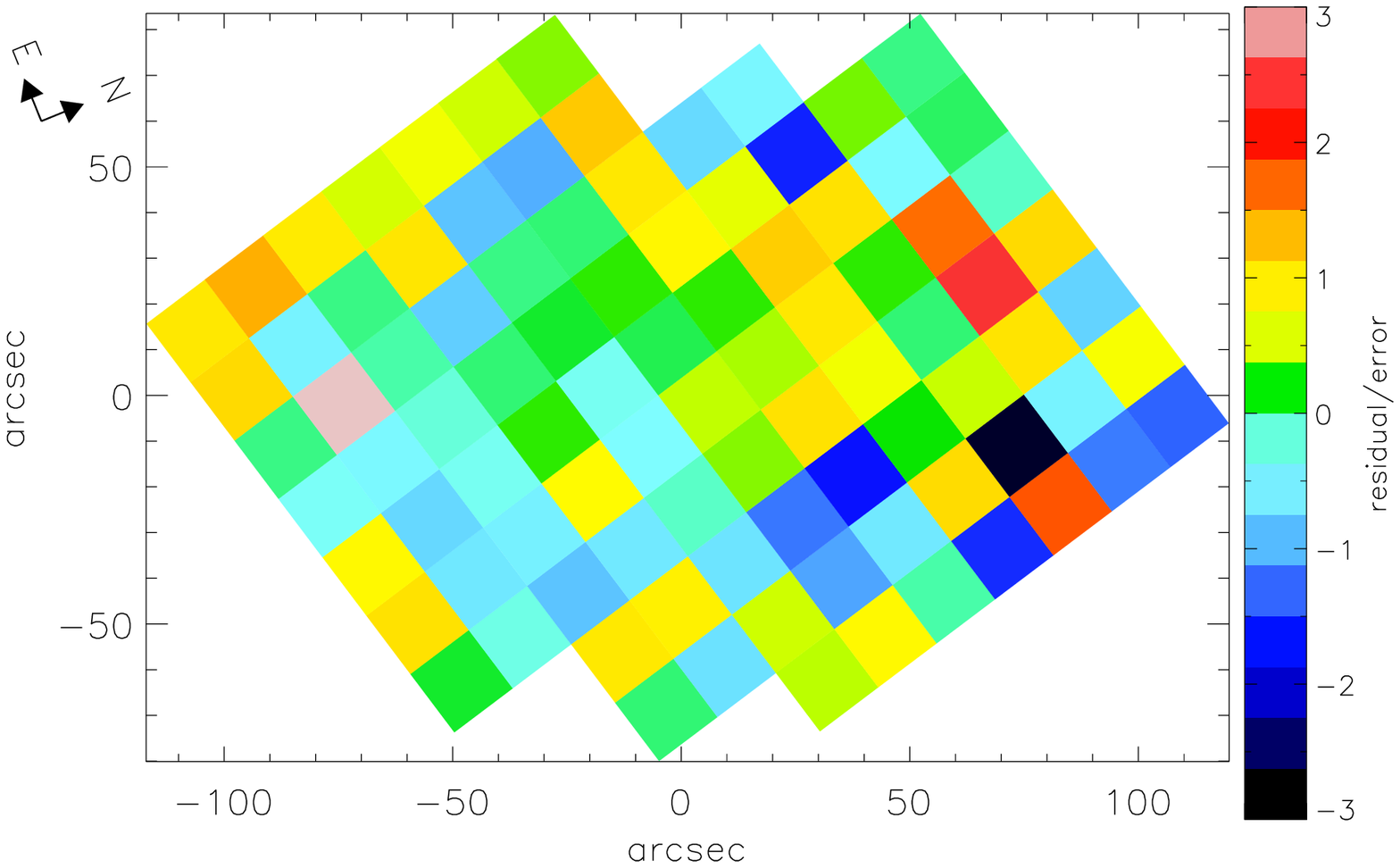}}\\
\subfigure{\includegraphics [scale=0.3,angle=0,trim=0 0.3in 0 1.18in,clip=true]{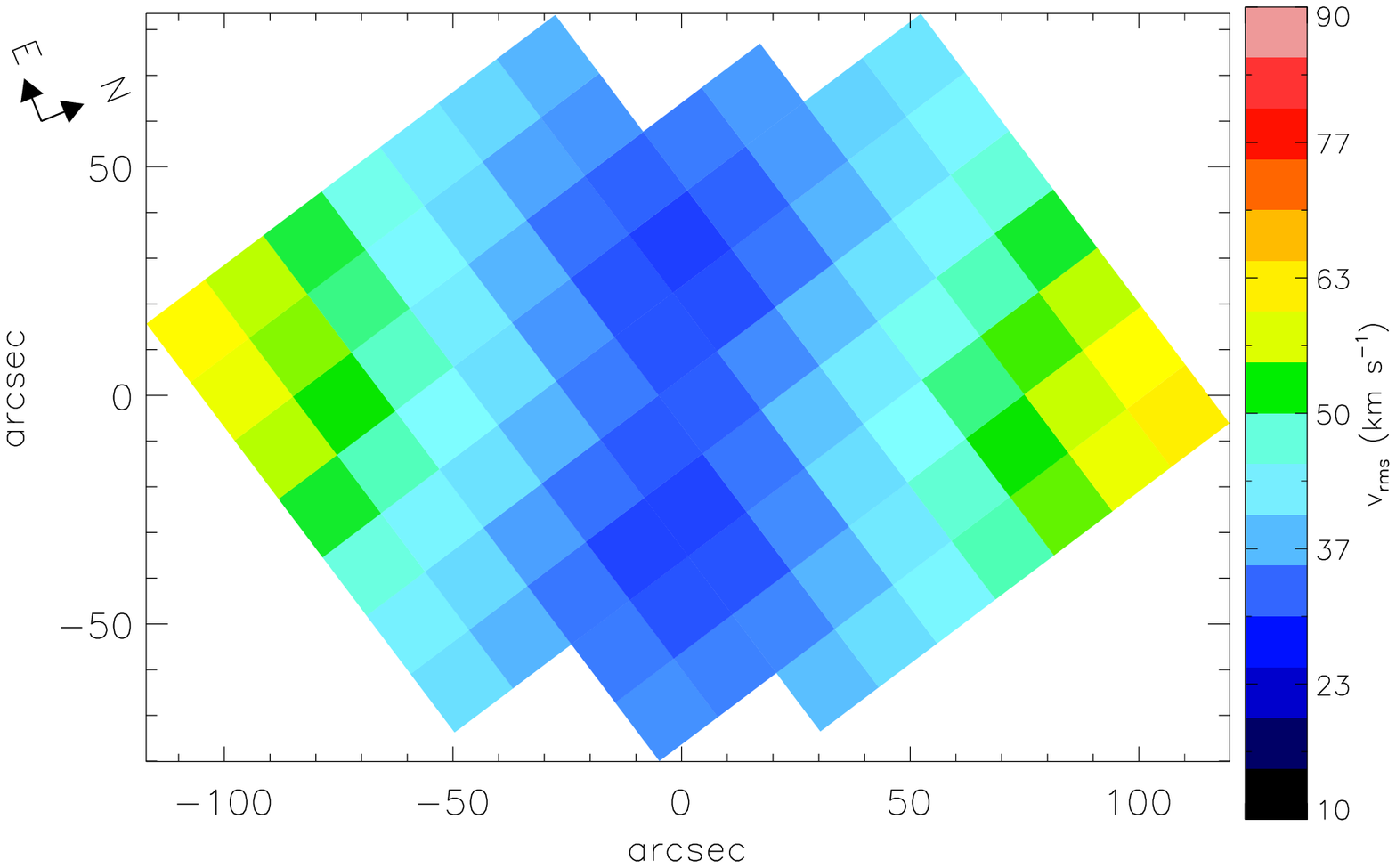}}
\subfigure{\includegraphics [scale=0.3,angle=0,trim=0 0.3in 0 1.18in,clip=true]{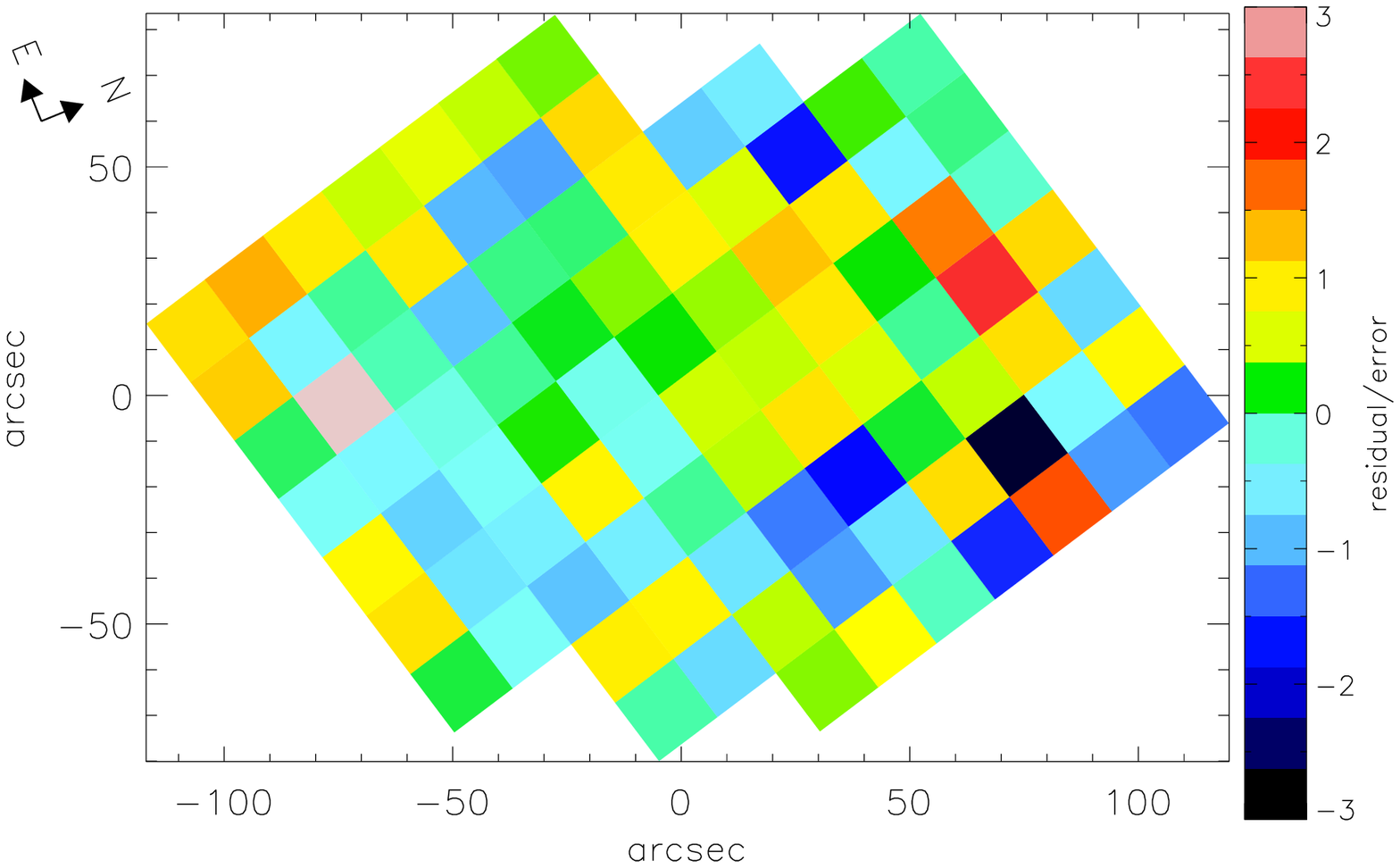}}\\
\subfigure{\includegraphics [scale=0.3,angle=0,trim=0 0.3in 0 1.18in,clip=true]{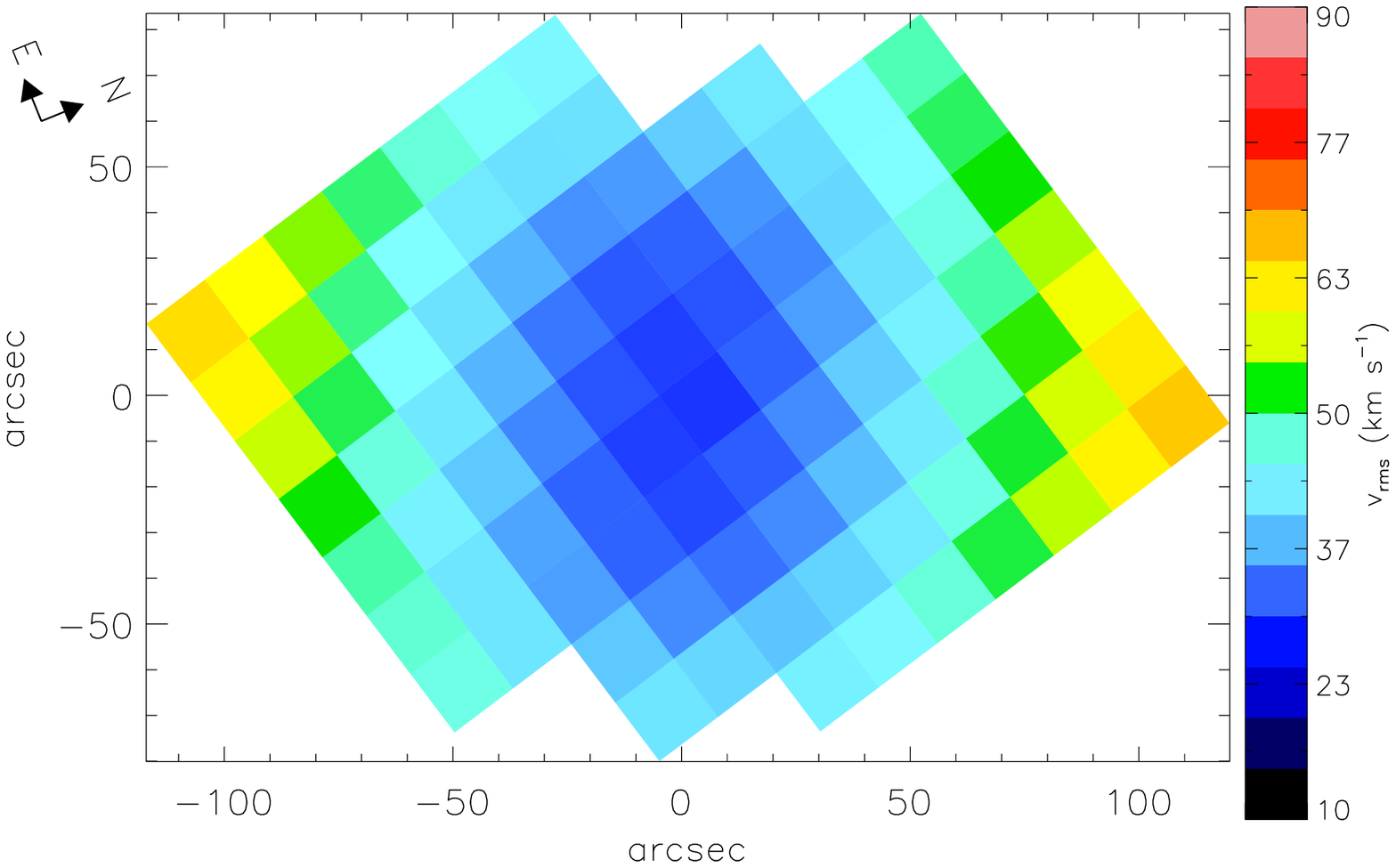}}
\subfigure{\includegraphics [scale=0.3,angle=0,trim=0 0.3in 0 1.18in,clip=true]{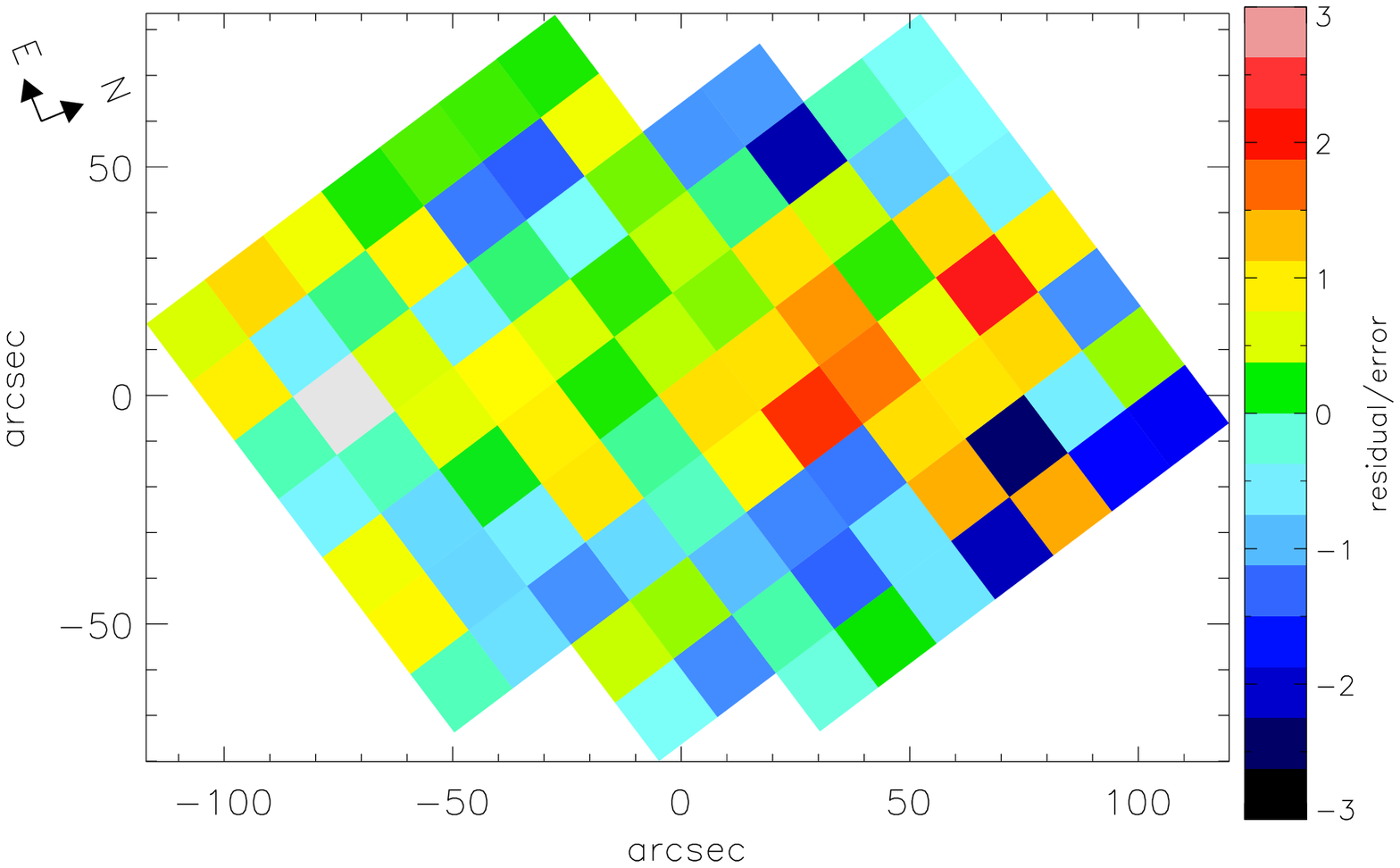}}
\caption{\textit{Left}: V$_{rms}$. \textit{Right}: V$_{rms}$ error 
for the data and residual/error for the models. 
\textit{Row 1}: Data. \textit{Row 2}: The model at 
the least-squares minimum fit made to the kinematic 
data alone. This model disagrees with the SPS determination on 
$\Upsilon_{*,R}$. The model is characterized by $\chi^2_{\nu=87}$ = 77.0.
\textit{Row 3}: The model at the least-squares minimum fit 
to the combined SPS and 
kinematic data. Including the 
SPS penalty, the fit is characterized by $\chi^2_{\nu=88}$ = 77.1. 
\textit{Row 4}: The model second-moment velocity map 
for a fit fixed to $\Upsilon_{*,R}$ = 1.1, $\alpha$ = 0.1, and the 
remaining parameters freely fit. This model represents the class of 
DM dominated and cored models that are excluded by our data.
The penalty in 
$\chi^2$ comes from model velocities that rise more slowly than 
the data along the 
major axis and excess model velocities at minor axis offsets. This 
fit is characterized by $\chi^2_{\nu=89}$ = 93.9.
}
\label{fig_JAM_model2}
\end{figure}

\begin{figure}
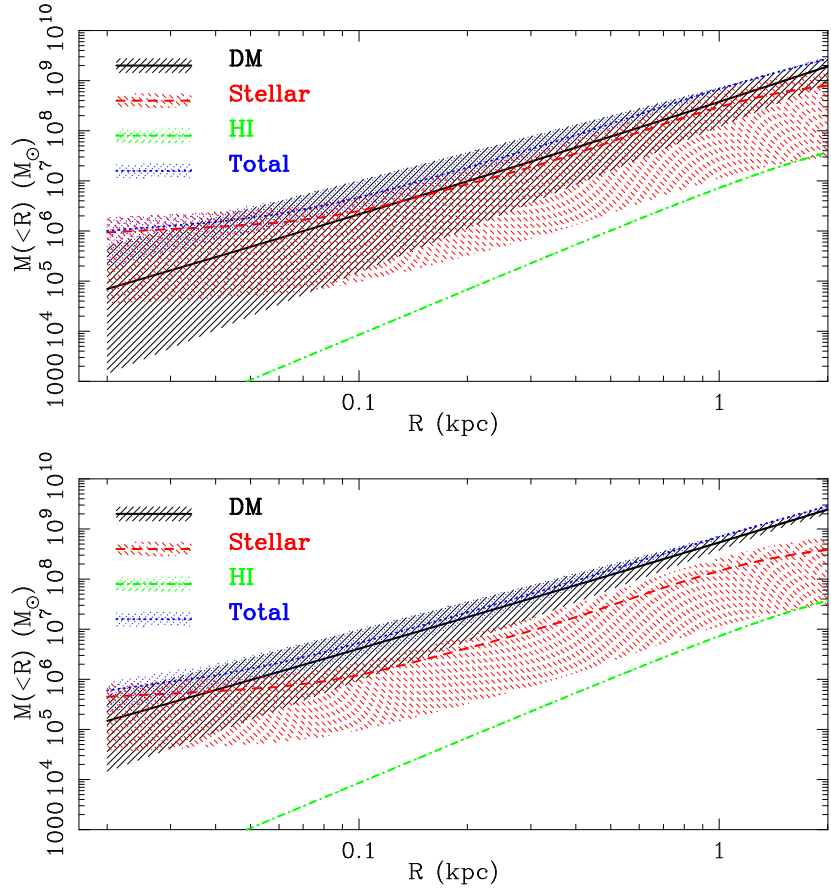

\centering
\subfigure{\includegraphics [scale=0.7,angle=-90]{pgplot_p5_v1.eps}}
\subfigure{\includegraphics [scale=0.7,angle=-90]{pgplot_p5_v2.eps}}
\caption{The enclosed mass profiles and confidence intervals. 
\textit{Top}: Using the kinematic data alone. 
The $\Upsilon_{*,R}$ fit here exceeds that fit through 
stellar population synthesis analysis. With these poor 
constraints, 
NGC~2976 may be DM or baryon dominated to the largest 
radius measured. The median models have similar amounts of 
enclosed DM and stellar mass up to 1 kpc where the 
DM begins to dominate.
\textit{Bottom}: Using the combined SPS and kinematic data. 
The DM halo dominates at least to r $>$ 200 pc and 
perhaps everywhere.}
\label{fig_mm4}
\end{figure}

The best-fit, five-parameter model has a shallow 
minimum at $\Upsilon_{*,R}$=3.49, 
$i=$63.3$\arcdeg$, $\beta_z=$0.432, $\rho_0(r=1 pc)=$0.260 M$_{\odot}$ 
pc$^{-3}$, and 
$\alpha=$0.235 at $\chi^2_{\nu=87}=$77.0. The second-moment model and residuals are 
shown in Figure \ref{fig_JAM_model2}. 
The enclosed mass distribution for this model is shown in 
Figure \ref{fig_mm4}. The anisotropy we measure is larger than the 
$\beta_z\sim0.3$ commonly fit by the same method in E's and 
SO's \citep{Capp07,Thom09} and slightly below the $\beta_z\sim0.5-0.8$ 
range commonly found in large spirals \citep{vdKr99,Shap03,Bers11}. 
We test an isotropic solution as well, and 
find a similar solution in the remaining parameters. 
The kinematically determined inclination is marginally above the 
value in \citet{deVa91} based on photometric ellipticity ($i=61.5$\arcdeg). However, we 
find $i=$64.6$\pm$1.5\arcdeg\ from our [OII] tilted ring [OII] fit \S \ref{sec_gasfit}, so 
the kinematically determined value is reasonable. Inclinations are often 
poorly constrained by kinematic fits \citep{Kraj05} and can have a 
strong degeneracy with $\beta_z$. There is tension 
in the $\Upsilon_*$ value compared to that which we determine from the 
stellar population synthesis (SPS) fits presented in \S \ref{sec_BC}. Using 
the constraint we derive from 
SPS, we add an additional $\chi^2$ term 
containing $\Delta\Upsilon_{*,R}$=0.8 with 
a central value of $\Upsilon_{*,R}$=1.1 to the $\chi^2$ values 
from the kinematics per
\begin{equation}
\label{eq_twochisq}
\chi^2=\left( \frac{\Upsilon_*-\hat{\Upsilon}_*}{\Delta\Upsilon_*}\right)^2+\sum_{i=1}^{N_{bins}}\left( \frac{v_{i,rms}-v_{i,model}}{\Delta v_{i,rms}}\right)^2.
\end{equation}

The parameters then reach a minimum at $\Upsilon_{*,R}$=1.158, 
$i=$65.0\arcdeg, $\beta_z=$0.450, $\rho_0(r=1 pc)=$45.7 M$_{\odot}$ 
pc$^{-3}$, and
$\alpha=$0.90 at $\chi^2_{\nu=88}=$77.1. By marginalizing over the 
other four parameters, the data disfavor $\alpha$=1 at 0.8$\sigma$ 
significance, $\alpha$=0.6 at 1$\sigma$
significance, and $\alpha$=0 at 2.2$\sigma$
significance.  
The $V_{rms}$ map and residuals are shown in Figure \ref{fig_JAM_model2}. 
The enclosed mass model is shown for the joint constraint in Figure \ref{fig_mm4}. 
This solution is within the 1$\sigma$ confidence interval from the 
kinematic data only. Finally, we fix $\Upsilon_{*,R}$=1 and 
$\alpha=$0.1 as an illustration of a cored DM model. The $V_{rms}$ map and residuals 
are shown in Figure \ref{fig_JAM_model2}. After optimizing the remaining three variables, this model 
yields $\chi^2_{\nu=88}=$93.9 and is excluded with high confidence. The 
crucial difference between this final model and the data is in the 
generally more circular V$_{rms}$ contour in the former. 

\label{sec_stelfit}
\subsection{Parameter degeneracies}
\label{sec_param_deg}
\begin{figure}
\centering
\subfigure{\includegraphics [scale=0.7,angle=0]{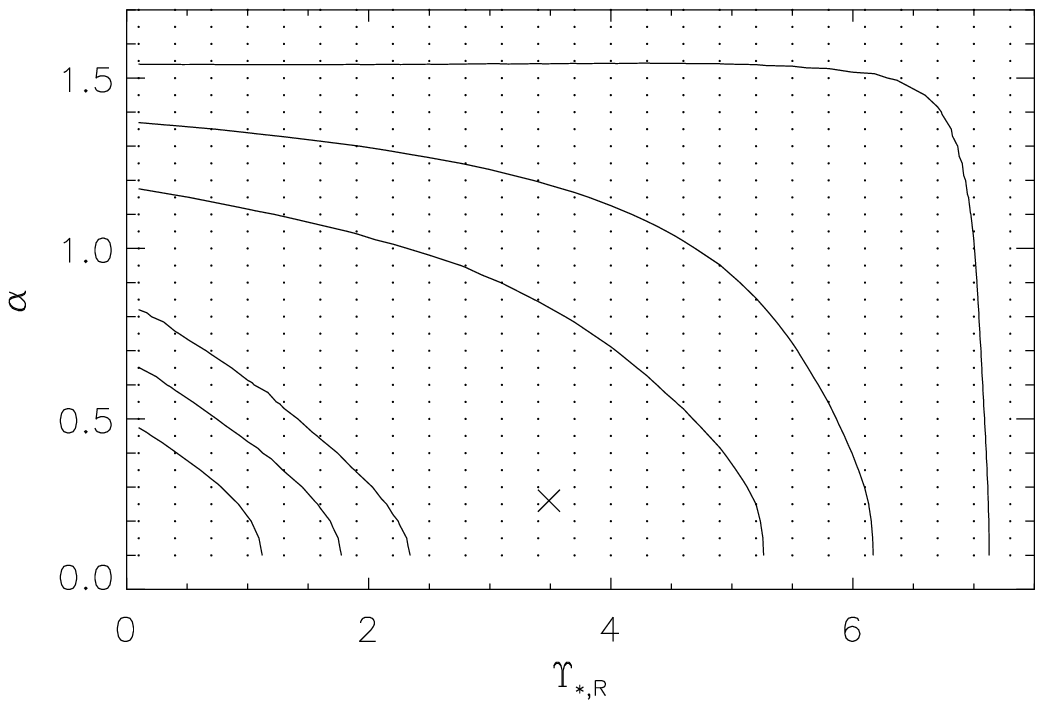}}
\subfigure{\includegraphics [scale=0.7,angle=0]{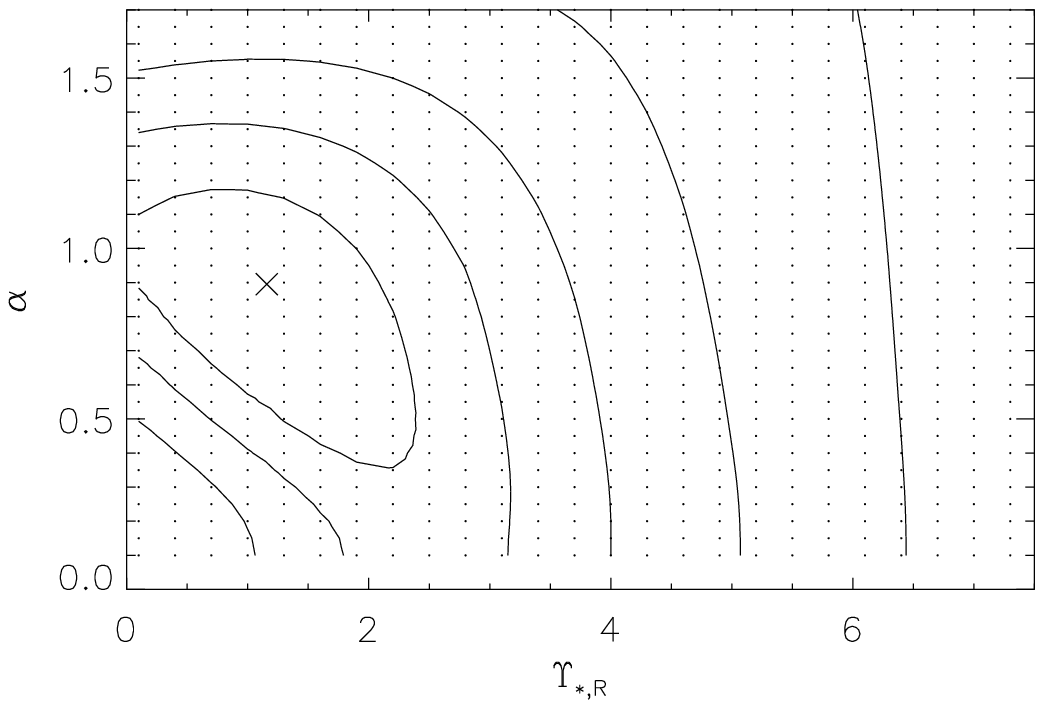}}
\caption{The $\Delta\chi^2$ distributions for combinations of the 
five parameters being minimized in the Jeans modelling. The 
inner three contours correspond to $\Delta\chi^2=$ 2.3, 
6.2, and 12.9 or 1, 2, and 3 $\sigma$ significance. The 
additional contours increase each by a factor of two. Cross symbols mark 
the formal minima. 
\textit{Left}: The confidence intervals from the 
kinematic data alone. \textit{Right}: The 
confidence intervals when the kinematic and 
SPS data are combined.}
\label{fig_param_deg}
\end{figure}

We here investigate the degeneracy between 
$\alpha$ and $\Upsilon_*$. By our parameterization, there is an obvious 
degeneracy between $\rho_0$ and $\alpha$ with a weaker 
degeneracy on $\Upsilon_*$ (Figure \ref{fig_param_deg}). 
The most important degeneracy for the purpose of constraining the 
mass budget is that betweeen $\Upsilon_*$ and $\alpha$. In 
NGC~2976, they anti-correlate. 
A similar exercise is done with the joint kinematic and SPS likelihoods 
by which 
the cored DM fit is excluded at 2$\sigma$ significance but a pure cusp 
is permitted. We conclude that honoring even a loose 
$\Upsilon_*$ limit makes a DM halo measurement 
entirely consistent with the NFW form. The cored model is 
disfavored at modest significance but amenable to stricter limits 
through more extended instrument pointings and higher S/N data.
\subsection{Other DM distributions}
\label{sec_ps_model}
We apply the power-law function as the primary 
DM density distribution, but we briefly test and discuss 
alternatives. The pseudo-isothermal function is commonly applied 
to DM halo data as a well-fitting cored model, 
although it lacks theoretical motivation. The pseudo-isothermal 
function 
\begin{equation}
\label{eq_pseudo}
\rho(r)=\frac{\rho_0}{1+(r/r_c)^2}
\end{equation}
contains a central 
DM space density of $\rho_0$ and scale length of $r_c$. In the 
limit of $r_c$ significantly larger than the datapoints, a power-law 
function with $\alpha=0$ mimics this function. However, the added 
flexibility of the scale length term can diminish the capability to 
discriminate between cores and cusps. The best-fit, five-parameter 
model with the SPS penalty has a minimum at $\Upsilon_{*,R}$=1.37, 
$i=$66.4$\arcdeg$, $\beta_z=$0.390, $\rho_0=$0.198 M$_{\odot}$ 
pc$^{-3}$, and $r_c=$1.0 kpc at $\chi^2_{\nu=88}=$78.3. The steep, 
power-law model is still preferred, but the 
cored model is statistically viable. Data at larger radii are necessary 
to better test for the presence of a large core.

Next, we test a six parameter model with the NFW form. 
A minimization of the full NFW function requires a large scale 
radius, $r_s$, for a quality fit. Only models with $r_s>10$ kpc 
fit well where the power-law approximation becomes highly precise. This 
results in an enormous DM virial mass ($\sim2\times10^{11} 
M_{\odot}$), but data at larger radii are necessary 
to make a reliable estimate of the virial mass.

\section{Models from gas kinematics}
\label{sec_gasfit}
\begin{figure}
\centering
\subfigure{\includegraphics [scale=0.4,angle=0]{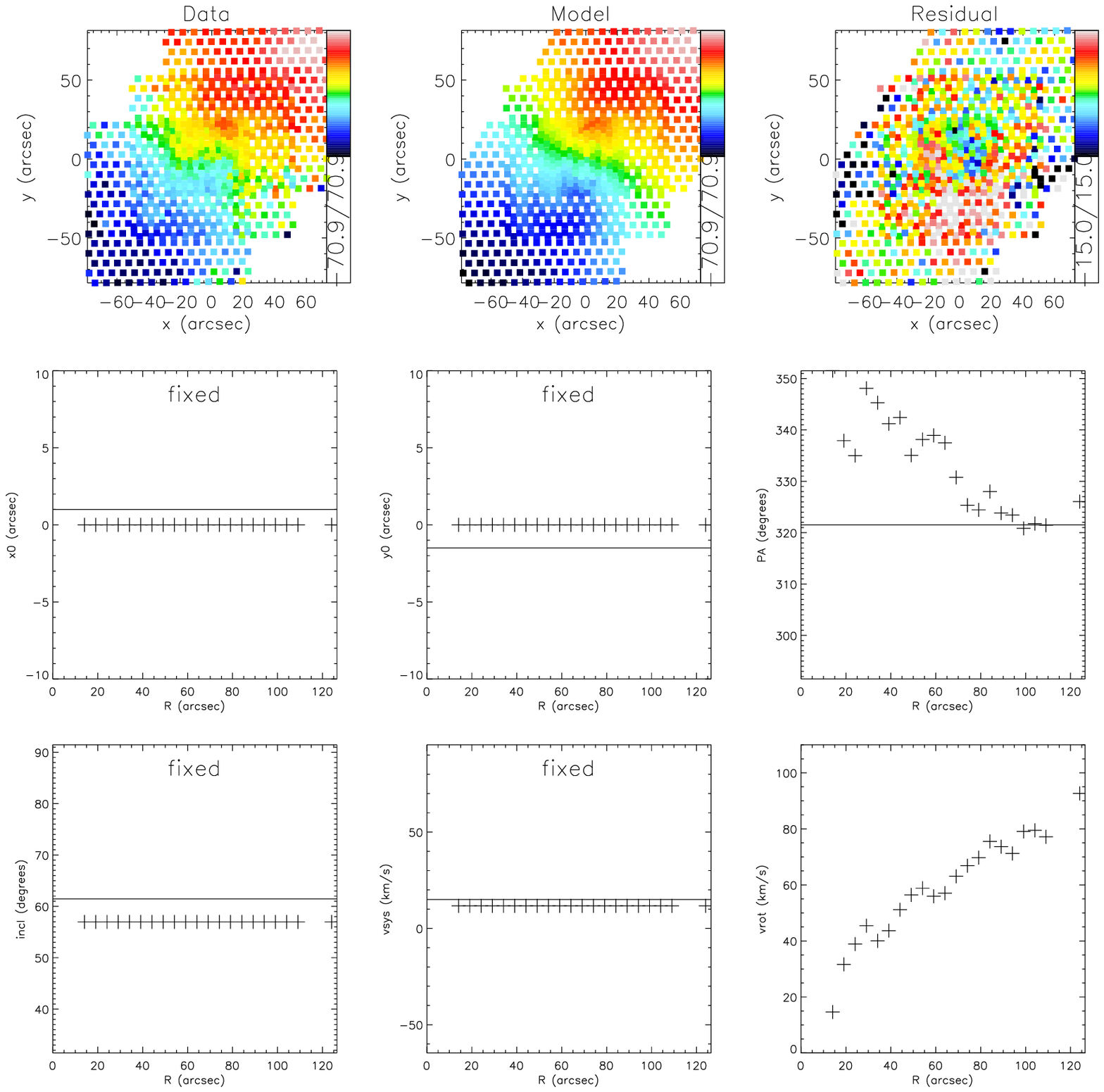}}
\subfigure{\includegraphics [scale=0.4,angle=0]{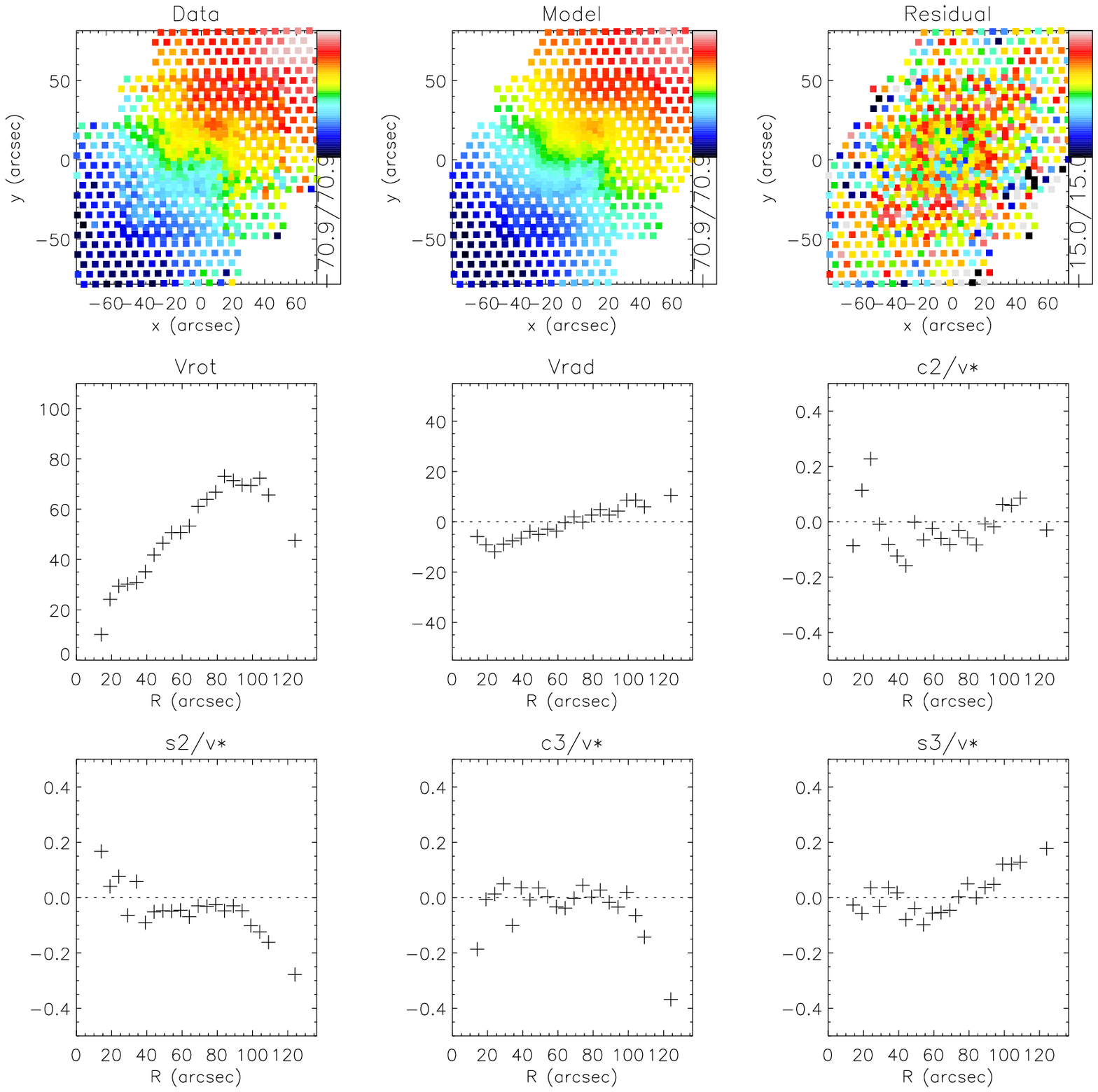}}
\caption{The data and fits to the [OII] velocity field 
representing the two ways that the kinematic twist can be recovered. 
\textit{Left}: The fit with a tilted ring (TR) 
model and a variable position angle. \textit{Right}: 
The fit with a harmonic decomposition (HD) model through three 
orders. Signficant radial velocity is observed, particularly 
in the galaxy's center. The sign of the radial velocity is 
uncertainty as we cannot identify the galaxy's near side. 
If we speculate that the galaxy's near side is in the SW, 
the gas near the center is outflowing. 
The higher-order terms are noisy, but 
similar rotational and radial velocity fits are made 
when the higher order terms are not fit.}
\label{fig_gas_hd}
\end{figure}

\begin{figure}
\centering
\includegraphics [scale=0.5,angle=-90]{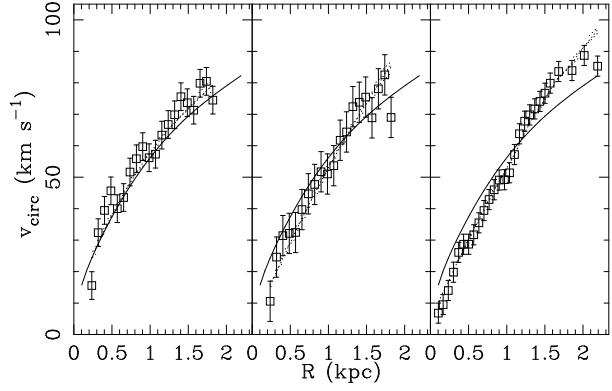}
\caption{Comparison between the circular velocity determinations 
from stellar and gas kinematics. The solid line in each panel shows the 
circular velocity for the mass model favored by the stellar kinematics 
with a SPS penalty (right panel of Figure \ref{fig_param_deg}). The 
circular velocity as determined by the gas kinematics is shown with 
datapoints and the dotted line as a power-law fit. The 68\% confidence 
interval is indicated by an upper and lower dotted line. The fits to the 
gas data use fixed $\Upsilon_{*}$ values per Table \ref{tab_gas_fit}. 
\textit{Left}: 
The [OII] data and fit for the tilted-ring model with a varying position angle. 
\textit{Middle}: The [OII] data and fit for the harmonic decomposition 
model through third order. The fit is steeper than the stellar 
based model and consistent with a cored DM halo. \textit{Right}: The 
H$\alpha$ data of SBLB03. These data are fit by removing a radial 
velocity component at small radius and similar to our [OII] harmonic 
decomposition model.}
\label{fig_gas_rot}
\end{figure}

The [OII] data are fit with a tilted ring (TR) \citep{Rogs74,Rogs76}
and harmonic decomposition (HD) 
algorithm to determine a rotation curve assuming an
infinitely thin geometry for the gas. The code is the 
same as used and described in \citet{Fath05}. The HD model 
and terms are shown in Figure \ref{fig_gas_hd} as is the 
TR model. Driven by the same
complex structures and kinematic twists as discussed
by SBLB03 for the H$\alpha$ in this galaxy,
we have make harmonic fits through the $m=3$ terms. SBLB03 
only present circular and radial terms, however. Our 
TR fit is allowed a position angle that varies with 
radius which can also explain the kinematic twist. 
The [OII] rotation
curves are given in Table \ref{tab_gas_rot} and
shown in Figure \ref{fig_gas_rot}. The 
asymmetric drift correction for the ionized gas in NGC~2976 
has been calculated in SBLB03, found to be small, and not 
used in their analysis because of substantial uncertainties in 
its exact value. Similarly, we do not apply an asymmetric drift 
correction to our gas rotation curve fits. There is 
consistency between the shapes of our [OII] rotation curve 
and the H$\alpha$ rotation curve of SBLB03. The irregular 
structure in the curve at r$=30$\arcsec\ and r$=60$\arcsec\ is 
found in both datasets, particularly in our tilted ring fit.

We next fit a velocity power
law of the form $v_{circ,DM}\propto r^{\beta}$ added in 
quadrature to the stellar and HI component circular velocities. 
The circular velocity of a power law density profile is a 
power law with a different index. 
The well-known relation for the density and circular velocity indices for power
laws of $\alpha=2\times(1-\beta)$ (e.g.~SBLB03 Appendix B) is used. 
A variety of $\Upsilon_*$ values 
are tested as detailed in Table \ref{tab_gas_fit}, some 
fixed and some fit live. We also refit the
SBLB03 data of their Table 3 in the same manner, taking their mass model
rotation curves and trying their maximal disk value of M$_*$/L$_K=0.19$. 
The residuals from the best fit were used to estimate the
uncertainty in the rotation curve; these 
systematic uncertainties are larger than the statistical errors and 
included in the error determination of $\alpha$. 

The estimates of $\alpha$ from the gas kinematics 
are presented in Table \ref{tab_gas_fit} for a variety of 
datasets and assumptions on $\Upsilon_*$. 
The HD fits to the [OII] data require a cored DM halo, regardless 
of the $\Upsilon_*$ assumptions. The [OII] TR fits require DM slopes that 
are steeper than the HD fits at 3$\sigma$ significance, 
but still deviating from 
NFW expectations at 1.5$\sigma$ significance. 
The harmonic decomposition can fit for 
radial infall or outflow, but some other motions may be 
degenerate with rotation such as the motions due to a 
bar-like potential \citep{Spek07}. A bar is expected to 
show power in the third order terms, and our m~=~3 sine term 
does show some power at large radius. 
Finally, we fit the harmonic decomposition data of SBLB03 in the same 
manner. From their fits with a range in M/L chosen to 
represent maximal and submaximal disks, SBLS03 
reach ranges of 0.01$<\alpha<$0.17. We find agreement with 
their determinations. We find a larger, but still deviating from 
a cusp by 5$\sigma$ significance, 
DM halo slope when we constrain the disk to have no mass. We conclude 
that a mass model based on the gas data with a single position angle and 
harmonic terms favors a cored DM halo, but that this result is not 
robust against an equally viable model that fits the data with a twist 
in the position angle. The mass profile conclusions drawn from the gas 
kinematics in this object are subject to severe dependencies in the 
modelling choices.

A comparison between the gas velocity field in Figure \ref{fig_gas_hd} 
and the stellar line-of-sight velocity in Figure \ref{fig_velmap} shows 
that the zero velocity contours in both are twisted in the same way and 
with similar magnitude along the minor axis. This may be an important 
clue to the cause of the non-regular motions as collisional processes 
are expected to not twist the stellar zero velocity contour. One explanation 
may be that both the stars and gas have their angular momentum vectors perturbed 
at small radius, perhaps by bending in a weak bar potential. In this case, 
the stellar models are more immune to a warp as the Jeans models compare 
to V$_{rms}$ instead of simply V$_{los}$, and we measure a stellar velocity 
dispersion that is larger than line-of-sight rotation out to R$_{maj}\sim$ 
60\arcsec. With the current level of stellar, observational errors, the 
axisymmetric models we present are statistically sufficient to describe the 
galaxy. Better data, taken in the future, may merit analysis with 
non-axisymmetric orbit-based models. 
\section{Conclusions}
\label{sec_concl}
We present two-dimensional maps of stellar and gaseous 
kinematics in the late-type 
dwarf galaxy NGC~2976. 
Theoretical models of collisionless, cold 
dark matter predict a cuspy dark matter halo to 
exist in low mass halos when gravity alone is simulated. Baryonic feedback 
processes have been proposed as a mechanism to create a cored halo, which 
observations such as ours may constrain. The leverage of the 
stellar kinematics as a collisionless tracer is a major 
advantage to our work. We fit the stellar kinematics with 
an axisymmetric, semi-isotropic Jeans model to measure the 
DM profile and constrain $\Upsilon_{*}$. The Jeans model 
permits both a cored or cuspy halo with a mild preference 
for a cored, high $\Upsilon_{*,R}$. We next fit a suite of 
stellar population histories to an optical spectrum 
with a broader bandpass. We find a limit of $\Upsilon_{*,R}$=1.1$\pm$0.8 
driven primarily by uncertainty in the initial mass function. This 
limit in combination with the kinematic data provides a much tighter 
certainty on the DM profile. The combined fit suggests a DM 
cusp ($\alpha=0.90\pm0.15$ at 1$\sigma$) and 
excludes a DM core at 2$\sigma$ significance. NGC~2976 is dark 
matter dominated 
everywhere outside of the nuclear star cluster and requires no 
history of baryonic feedback or non-standard particle properties 
to explain the dark matter halo profile. 
The gaseous kinematics, in concordance with earlier 
work, imply a cored dark matter halo when modeled with a 
harmonic decomposition method and a single position angle. A tilted-ring 
fit with a variable position angle instead excludes the cored model 
and is compatible with the models fit by the stellar kinematics. 
A larger sample with stellar kinematics is motivated to compare with the vast 
literature on gaseous kinematics and reassess the quantity and distribution 
of dark matter and possible correlations with central density slope 
in late-type dwarf galaxies. This work comes with two primary caveats 
that can be improved upon. First, our results are strongest 
for a DM density power-law approximation, and a DM density model with 
a large core under a different parameterization 
cannot be strictly excluded with the current data as 
discussed in \S \ref{sec_ps_model}. Data at larger radii, such as 
$>$2 kpc, will close this uncertainty but require a deeper flux limit. 
Second, a crucial barrier to achieving lower 
statistical errors in this work has been the large instrumental resolution 
relative to the intrinsic dispersion. In future works, we will 
use the newly available Visible Integral field Replicable Unit 
Spectrograph Wendelstein (VIRUS-W) \citep{Fabr08} 
at a resolution of $R=6800$ to circumvent this limitation.
\acknowledgments
JJA acknowledges a National Science 
Foundation Graduate Student Fellowship and a UT Harrington Endowment 
Dissertation Fellowship that have supported him 
through this work. KG acknowledges support from 
NSF-0908639. We are grateful to Josh Simon for 
providing important suggestions during this paper's editing. 
We thank George and Cynthia Mitchell for funding the VIRUS-P instrument. 
This research has made use of the NASA/IPAC Extragalactic Database (NED) 
which is operated by the Jet Propulsion Laboratory, California Institute 
of Technology, under contract with the National Aeronautics and 
Space Administration. We have used MGE and JAM 
as written and distributed by Michele Cappellari; we 
thank him for his fine work. This work made use of 
THINGS, `The HI Nearby Galaxy Survey', and 
SINGS, the `Spitzer Infrared Nearby Galaxies Survey'; we thank them for their public 
data release. We note that the public availability of the \citet{Simo03} CO data and 
tabular rotations curves were used here for comparison and appreciated. 
Finally, we thank an anonymous referee for important improvements to 
this work. 
{\it Facilities:} \facility{Smith (VIRUS-P)}.

\bibliography{NGC2976_DM}   
\bibliographystyle{apj}   

\clearpage

\clearpage

\begin{deluxetable}{lrrrrr}
\tabletypesize{\scriptsize}
\tablecaption{Kinematic template stars\label{tab_kintemp}}
\tablewidth{0pt}
\tablehead{
\colhead{HYPERLEDA \#} & \colhead{HD \#} & \colhead{Type} & \colhead{[Fe/H]} & \colhead{Average weight}
& \colhead{Intrinsic $\sigma$ (km s$^{-1}$)}}
\startdata
00672 & HD181214 & F8III & -0.01 & 0.086 & 30.8 \\
01269 & HD058923 & F0III &  0.21 & 0.185 & 0 \\
01286 & HD062509 & K0IIIb &  0.02 & 0.006 & 11.1 \\
01298 & HD068017 & G4V &  -0.41 & 0.067 & 10.6 \\
01322 & HD073667 & K1V &  -0.55 & 0.032 & 10.9 \\
01359 & HD088609 & G5IIIwe &  -2.67 & 0.197 & 12.3 \\
01366 & HD089744 & F7V &  0.11 & 0.008 & 15.3 \\
01733 & HD169985 & G0III+ &  0.34 & 0.138 & 12.3 \\
01828 & HD193793 & WC+ &  0.77 & 0.280 & 0 \\
\enddata
\end{deluxetable}

\LongTables
\begin{deluxetable}{lrrrrrr}
\tabletypesize{\scriptsize}
\tablecaption{Stellar kinematic measurements in NGC~2976\tablenotemark{$\ast$}\label{tab_stelkin}}
\tablewidth{0pt}
\tablehead{
\colhead{Bin \#\tablenotemark{$\dagger$}} & \colhead{Major axis} & \colhead{Minor axis} &\colhead{v$_{los,*}$}
& \colhead{v$_{los,*}$, 1 $\sigma$} & \colhead{$\sigma_{*}$} & \colhead{$\sigma_{*}$, 1 $\sigma$} \\
& \colhead{distance} & \colhead{distance} &
& \colhead{uncertainty} & & \colhead{uncertainty} \\
& \colhead{(\arcsec)} & \colhead{(\arcsec)} &\colhead{(km s$^{-1}$)}
& \colhead{(km s$^{-1}$)} & \colhead{(km s$^{-1}$)} & \colhead{(km s$^{-1}$)}}
\startdata
1 & -48.1 & -62.5 & -34.53 & 26.29 & 10.00 & 49.63 \\
2 & -57.7 & -49.7 & -55.44 & 15.52 & 10.00 & 34.29 \\
3 & -67.3 & -37.0 & -34.53 & 18.81 & 48.27 & 24.73 \\
4 & -76.9 & -24.2 & -45.49 &  3.90 & 10.00 & 18.03 \\
5 & -86.5 & -11.4 & -51.80 &  5.01 & 10.00 & 19.68 \\
6 & -96.2 &  1.3 & -62.00 &  4.67 & 10.00 & 19.78 \\
7 & -105.8 & 14.1 & -65.17 &  7.47 & 10.00 & 23.18 \\
8 & -35.3 & -52.9 & 27.51 &  9.08 & 10.00 & 26.72 \\
9 & -44.9 & -40.1 & -11.36 &  5.43 & 24.54 & 12.43 \\
10 & -54.5 & -27.3 & -14.32 &  4.37 & 30.58 &  7.89 \\
11 & -64.2 & -14.6 & -32.05 &  3.31 & 29.10 &  6.55 \\
12 & -73.8 & -1.8 & -43.70 &  3.18 & 46.88 &  5.49 \\
13 & -83.4 & 11.0 & -42.71 &  4.22 & 22.76 & 10.50 \\
14 & -93.0 & 23.8 & -45.30 & 10.15 & 56.69 & 17.22 \\
15 & -3.3 & -68.8 & -26.18 & 15.56 & 10.00 & 32.14 \\
16 & -12.9 & -56.0 & -6.00 &  9.61 & 39.36 & 11.88 \\
17 & -22.5 & -43.2 & -1.24 &  5.14 & 16.88 & 12.89 \\
18 & -32.1 & -30.5 & -9.45 &  2.73 & 26.91 &  7.21 \\
19 & -41.8 & -17.7 & -18.39 &  1.71 & 30.71 &  5.34 \\
20 & -51.4 & -4.9 & -26.85 &  1.43 & 31.07 &  4.85 \\
21 & -61.0 &  7.9 & -29.12 &  1.50 & 34.13 &  4.89 \\
22 & -70.6 & 20.6 & -25.46 &  3.28 & 39.24 &  5.65 \\
23 & -80.3 & 33.4 & -39.14 &  9.02 & 43.85 & 13.96 \\
24 &  9.5 & -59.1 & -11.57 &  6.45 & 15.23 & 13.59 \\
25 & -0.1 & -46.4 & -12.83 &  4.69 & 31.38 &  7.91 \\
26 & -9.7 & -33.6 &  1.99 &  2.61 & 22.52 &  7.57 \\
27 & -19.4 & -20.8 & -8.08 &  1.64 & 32.75 &  4.94 \\
28 & -29.0 & -8.0 & -12.27 &  1.40 & 32.78 &  4.62 \\
29 & -38.6 &  4.7 & -14.64 &  1.35 & 34.33 &  4.71 \\
30 & -48.3 & 17.5 & -12.42 &  1.73 & 32.77 &  5.03 \\
31 & -57.9 & 30.3 & -9.70 &  3.21 & 45.44 &  5.29 \\
32 & -67.5 & 43.1 & -7.36 &  8.60 & 48.44 & 14.59 \\
33 & 31.9 & -62.3 & -11.61 &  5.59 & 35.13 & 10.25 \\
34 & 22.3 & -49.5 & -1.60 &  3.29 & 33.96 &  7.72 \\
35 & 12.7 & -36.7 &  1.96 &  2.54 & 22.54 &  7.65 \\
36 &  3.0 & -24.0 &  1.11 &  1.85 & 26.62 &  5.67 \\
37 & -6.6 & -11.2 & -6.59 &  1.73 & 25.43 &  5.62 \\
38 & -16.2 &  1.6 & -7.58 &  1.54 & 27.82 &  5.04 \\
39 & -25.9 & 14.4 & -4.36 &  1.32 & 32.51 &  4.59 \\
40 & -35.5 & 27.2 & -7.92 &  1.84 & 32.87 &  4.89 \\
41 & -45.1 & 39.9 & -10.69 &  3.90 & 25.57 &  8.49 \\
42 & -54.7 & 52.7 & -26.01 &  8.98 & 38.92 & 16.60 \\
43 & 44.7 & -52.7 &  9.80 &  5.35 & 41.22 &  9.30 \\
44 & 35.1 & -39.9 &  7.40 &  3.25 & 24.71 &  7.73 \\
45 & 25.4 & -27.1 &  3.85 &  1.69 & 23.17 &  6.58 \\
46 & 15.8 & -14.3 &  5.60 &  1.86 & 31.52 &  4.77 \\
47 &  6.2 & -1.5 &  6.45 &  1.43 & 30.24 &  4.65 \\
48 & -3.5 & 11.2 &  5.29 &  1.50 & 28.17 &  4.99 \\
49 & -13.1 & 24.0 &  7.00 &  1.74 & 28.55 &  5.06 \\
50 & -22.7 & 36.8 &  5.31 &  1.93 & 29.17 &  6.11 \\
51 & -32.4 & 49.6 & -2.15 &  5.02 & 18.75 & 12.36 \\
52 & -42.0 & 62.4 & -7.00 & 19.10 & 45.28 & 28.29 \\
53 & 57.5 & -43.0 & 11.07 &  3.43 & 38.57 &  6.14 \\
54 & 47.8 & -30.2 & 15.94 &  2.74 & 31.43 &  6.30 \\
55 & 38.2 & -17.5 & 13.44 &  1.52 & 25.29 &  5.70 \\
56 & 28.6 & -4.7 & 16.54 &  1.54 & 34.99 &  4.28 \\
57 & 18.9 &  8.1 & 20.57 &  1.43 & 26.04 &  5.32 \\
58 &  9.3 & 20.9 & 14.82 &  1.53 & 25.41 &  5.78 \\
59 & -0.3 & 33.7 &  7.83 &  2.53 & 29.96 &  5.51 \\
60 & -10.0 & 46.4 & -1.65 &  4.03 & 34.12 &  6.17 \\
61 & -19.6 & 59.2 & 34.47 &  7.82 & 27.31 & 17.13 \\
62 & -29.2 & 72.0 & 43.28 & 66.67 & 10.00 & 44.09 \\
63 & 70.2 & -33.4 & 25.36 &  3.12 & 27.98 &  6.72 \\
64 & 60.6 & -20.6 & 25.45 &  2.55 & 42.12 &  4.86 \\
65 & 51.0 & -7.8 & 22.35 &  1.68 & 36.79 &  4.00 \\
66 & 41.3 &  5.0 & 19.14 &  1.46 & 36.58 &  4.02 \\
67 & 31.7 & 17.7 & 18.32 &  1.47 & 33.43 &  4.37 \\
68 & 22.1 & 30.5 &  9.92 &  1.62 & 34.08 &  4.60 \\
69 & 12.4 & 43.3 &  4.18 &  3.05 & 31.55 &  5.96 \\
70 &  2.8 & 56.1 & -5.93 &  6.36 & 10.00 & 18.74 \\
71 & 83.0 & -23.8 & 44.14 &  3.44 & 39.98 &  6.14 \\
72 & 73.4 & -11.0 & 37.15 &  2.95 & 10.00 & 14.62 \\
73 & 63.7 &  1.8 & 32.91 &  2.67 & 36.58 &  6.40 \\
74 & 54.1 & 14.6 & 32.88 &  1.79 & 27.56 &  5.30 \\
75 & 44.5 & 27.4 & 20.85 &  2.07 & 32.52 &  4.91 \\
76 & 34.8 & 40.2 &  7.99 &  2.92 & 37.76 &  5.12 \\
77 & 25.2 & 53.0 &  5.11 &  4.08 & 10.00 & 15.89 \\
78 & 15.6 & 65.7 & -7.83 &  6.33 & 23.54 & 11.61 \\
79 & 95.8 & -14.1 & 44.77 &  3.09 & 24.18 &  7.57 \\
80 & 86.2 & -1.3 & 44.70 &  2.81 & 26.93 &  6.62 \\
81 & 76.5 & 11.4 & 37.63 &  3.08 & 39.61 &  4.68 \\
82 & 66.9 & 24.2 & 31.54 &  2.88 & 45.35 &  4.49 \\
83 & 57.3 & 37.0 & 31.35 &  4.23 & 38.50 &  6.47 \\
84 & 47.6 & 49.8 & 19.65 &  7.23 & 25.00 & 13.79 \\
85 & 38.0 & 62.6 & -5.33 & 11.58 & 37.91 & 17.95 \\
86 & 108.6 & -4.5 & 38.98 &  7.02 & 27.43 & 11.96 \\
87 & 98.9 &  8.3 & 43.17 &  7.08 & 44.62 &  9.69 \\
88 & 89.3 & 21.1 & 22.96 &  5.53 & 40.10 &  8.91 \\
89 & 79.7 & 33.9 & 40.93 & 12.32 & 48.21 & 16.14 \\
90 & 70.0 & 46.7 &  2.96 &  8.51 & 39.16 & 17.83 \\
91 & 60.4 & 59.4 & -16.52 &  8.01 & 35.10 & 19.13 \\
92 & 50.7 & 72.2 & 20.40 & 40.00 & 25.82 & 24.16 \\

\enddata
\tablenotetext{$\ast$}{Column 4 contains the systemic velocity which
we measure as a weighted average to be 4.60 km s$^{-1}$. The quantity
fit in the Jeans models is the velocity second-moment (V$_{rms}$), or the sum in quadrature
of columns 4 and 6 with the systematic velocity removed.
The V$_{rms}$ error is formed by the propagation
of the listed terms as well as an estimated 0.5 km s$^{-1}$ uncertainty on the
average 50 km s$^{-1}$ instrumental resolution.}
\tablenotetext{$\dagger$}{The central position of each bin is given. The
bins are 16\arcsec$\times$16\arcsec in size and rotated by -53\arcdeg\ east of north.}
\end{deluxetable}

\begin{deluxetable}{lrrrrr}
\tabletypesize{\scriptsize}
\tablecaption{Rotation curve data\label{tab_gas_rot}}
\tablewidth{0pt}
\tablehead{
\colhead{Radius} & \colhead{v$_{c,TR,[OII]}$\tablenotemark{$\dagger$}} & \colhead{v$_{c,HD,[OII]}$\tablenotemark{$\dagger$}} & \colhead{v$_{rad,[OII]}$} & \colhead{v$_{c,*}$\tablenotemark{$\ast$}} & \colhead{v$_{c,HI}$} \\
\colhead{(arcsec)} & \colhead{(km s$^{-1}$)} & \colhead{(km s$^{-1}$)} & \colhead{(km s$^{-1}$)} & \colhead{(km s$^{-1}$)} & \colhead{(km s$^{-1}$)}}
\startdata
14.1 & 15.6 $\pm$ 0.8 & 10.6 $\pm$ 0.8 & 5.2 $\pm$ 0.4 & 9.5 & 1.5 \\
19.1 & 32.4 $\pm$ 0.5 & 24.6 $\pm$ 1.0 & 9.0 $\pm$ 0.6 & 11.2 & 2.0 \\
24.1 & 39.4 $\pm$ 0.9 & 31.4 $\pm$ 2.5 & 11.0 $\pm$ 0.7 & 13.0 & 2.5 \\
29.1 & 45.7 $\pm$ 0.8 & 32.1 $\pm$ 1.6 & 6.4 $\pm$ 1.5 & 15.0 & 3.0 \\
34.1 & 40.0 $\pm$ 0.4 & 32.4 $\pm$ 5.9 & 6.7 $\pm$ 1.6 & 16.9 & 3.5 \\
39.1 & 43.5 $\pm$ 1.1 & 39.7 $\pm$ 2.9 & 7.4 $\pm$ 1.8 & 18.8 & 3.9 \\
44.1 & 51.6 $\pm$ 0.6 & 44.7 $\pm$ 3.3 & 2.0 $\pm$ 1.8 & 20.5 & 4.4 \\
49.1 & 55.9 $\pm$ 0.7 & 47.7 $\pm$ 3.3 & 3.8 $\pm$ 1.3 & 22.1 & 4.9 \\
54.1 & 59.7 $\pm$ 1.0 & 51.7 $\pm$ 0.8 & 3.3 $\pm$ 1.0 & 23.5 & 5.3 \\
59.1 & 56.2 $\pm$ 0.5 & 51.0 $\pm$ 2.8 & 2.5 $\pm$ 5.5 & 24.8 & 5.7 \\
64.1 & 57.3 $\pm$ 0.2 & 53.7 $\pm$ 0.4 & -0.3 $\pm$ 0.9 & 25.9 & 6.1 \\
69.1 & 63.4 $\pm$ 0.3 & 61.8 $\pm$ 0.8 & -2.2 $\pm$ 2.1 & 26.8 & 6.5 \\
74.1 & 66.7 $\pm$ 0.4 & 64.4 $\pm$ 0.8 & 0.1 $\pm$ 1.1 & 27.6 & 6.9 \\
79.1 & 69.9 $\pm$ 0.2 & 72.4 $\pm$ 4.8 & -4.0 $\pm$ 2.3 & 28.2 & 7.2 \\
84.1 & 75.6 $\pm$ 0.5 & 73.8 $\pm$ 0.9 & -3.5 $\pm$ 2.7 & 28.7 & 7.6 \\
89.1 & 73.7 $\pm$ 0.5 & 75.5 $\pm$ 3.2 & -6.7 $\pm$ 5.3 & 29.0 & 7.9 \\
94.1 & 71.3 $\pm$ 0.6 & 68.9 $\pm$ 1.6 & -4.2 $\pm$ 2.1 & 29.2 & 8.2 \\
99.1 & 79.9 $\pm$ 0.5 & 78.1 $\pm$ 13.8 & -17.2 $\pm$ 10.0 & 29.4 & 8.5 \\
104.1 & 80.5 $\pm$ 0.9 & 82.5 $\pm$ 8.4 & -14.8 $\pm$ 7.7 & 29.4 & 8.7 \\
109.1 & 74.5 $\pm$ 2.1 & 69.0 $\pm$ 10.9 & -10.8 $\pm$ 3.8 & 29.3 & 9.0
\enddata
\tablenotetext{$\dagger$}{The listed errors are statistical. The
rms velocities from the best-fit TR and HD models are 4.5 km s$^{-1}$ and 6.1 km s$^{-1}$
for $\Upsilon_{*,R}=0$, 4.4 km s$^{-1}$ and 6.4 km s$^{-1}$
for $\Upsilon_{*,R}=1.1$, and 4.5 km s$^{-1}$ and 6.0 km s$^{-1}$
for a freely fit $\Upsilon_{*,R}$, respectively.}
\tablenotetext{$\ast$}{Assuming $\Upsilon_{*,R}=1$. This
column scales as $\propto$ ($\Upsilon_{*,R}$)$^{1/2}$}
\end{deluxetable}

\begin{deluxetable}{crrrr}
\tabletypesize{\scriptsize}
\tablecaption{Multi-Gaussian Expansion terms for NGC~2976\label{tab_MGE}}
\tablewidth{0pt}
\tablehead{
\colhead{Component} & \colhead{Index} & \colhead{$\Sigma_{0,k}$} & \colhead{$\sigma_k$} & \colhead{q$'_k$} \\
& \colhead{(k)} & \colhead{(M$_{\odot}$ pc$^{-2}$)} & \colhead{(arcsec)} & \tablenotemark{$\ast$}}
\startdata
Stellar\tablenotemark{$\dagger$} & 1 & 2298.3 & 0.28 & 1.00 \\
& 2 & 68.9 & 0.99 & 1.00 \\
& 3 & 27.8 & 6.36 & 1.00 \\
& 4 & 117.8 & 49.89 & 0.48 \\
& 5 & 19.6 & 108.26 & 0.54 \\
\hline
HI & 1 & 8.0 & 84.80 & 0.70 \\
& 2 & 1.3 & 291.13 & 0.75
\enddata
\tablenotetext{$\dagger$}{R-band and assuming $\Upsilon_*$=1 for
this listing.}
\tablenotetext{$\ast$}{Observed axial ratio.}
\end{deluxetable}


\begin{deluxetable}{lrrr}
\tabletypesize{\scriptsize}
\tablecaption{Dark matter density index constraints from gaseous kinematics\tablenotemark{$\dagger$}\label{tab_gas_fit}}
\tablewidth{0pt}
\tablehead{
\colhead{Assumption} & \colhead{This dataset's} & \colhead{This dataset's} & \colhead{SBLB03 dataset} \\
& \colhead{tilted ring fit} & \colhead{harmonic decomposition} &}
\startdata
$\Upsilon_{*}=$0 & 0.80 $\pm$ 0.08 & 0.31 $\pm$ 0.13 & 0.43 $\pm$ 0.06 \\
$\Upsilon_{*}$ fixed\tablenotemark{$\dagger\dagger$} & 0.84 $\pm$ 0.10 & 0.14 $\pm$ 0.17 & 0.16 $\pm$ 0.08 \\
freely fit $\Upsilon_{*}$ & 0.81 $\pm$ 2.20 & 0.31 $\pm$ 0.98 & 0.12 $\pm$ 0.44 \\
\enddata
\tablenotetext{$\dagger$}{The confidence ranges are based on rms circular velocity uncertainties
in order to capture both statistical and systematic errors.}
\tablenotetext{$\dagger\dagger$}{Assuming $\Upsilon_{*,R}=$ 1.1 for
our data and $\Upsilon_{*,K}=$ 0.19 for
the SBLB03 data.}
\end{deluxetable}

\end{document}